\documentclass[11pt]{article}

\usepackage{fullpage}
\usepackage{setspace}
\usepackage{parskip}
\usepackage{titlesec}
\usepackage[section]{placeins}
\usepackage{xcolor,colortbl} 
\usepackage{breakcites}
\usepackage{lineno}

\usepackage[utf8]{inputenc}
\usepackage[english]{babel}

\usepackage{subcaption}
\usepackage{float}
\usepackage{natbib}

\usepackage{graphicx}
\usepackage{multirow}
\usepackage{amsmath,amssymb,amsfonts}
\usepackage{amsthm}
\usepackage{mathrsfs}
\usepackage[title]{appendix}
\usepackage{xcolor}
\usepackage{textcomp}
\usepackage{manyfoot}
\usepackage{booktabs}
\usepackage{algorithm}
\usepackage{algorithmicx}
\usepackage{algpseudocode}
\usepackage{listings}

\usepackage[space]{grffile}
\usepackage{latexsym}
\usepackage{longtable}
\usepackage{tabulary}
\usepackage{array}
\usepackage{tabularx}

\usepackage{times}

\usepackage{soul}
\definecolor{ffffff}{HTML}{ffffff}
\definecolor{ff0000}{HTML}{ff0000}
\definecolor{ff0b0b}{HTML}{ff0b0b}
\definecolor{fff8f8}{HTML}{fff8f8}
\definecolor{fffcfc}{HTML}{fffcfc}
\definecolor{fffdfd}{HTML}{fffdfd}
\definecolor{ffd7d7}{HTML}{ffd7d7}
\definecolor{ff0606}{HTML}{ff0606}
\definecolor{fffefe}{HTML}{fffefe}
\definecolor{ff1212}{HTML}{ff1212}
\definecolor{ffdfdf}{HTML}{ffdfdf}
\definecolor{fff4f4}{HTML}{fff4f4}
\definecolor{ff6767}{HTML}{ff6767}
\definecolor{ffeaea}{HTML}{ffeaea}
\definecolor{ffdddd}{HTML}{ffdddd}
\definecolor{ffeeee}{HTML}{ffeeee}
\definecolor{ffe8e8}{HTML}{ffe8e8}
\definecolor{ffdddd}{HTML}{ffdddd}
\definecolor{ff6b6b}{HTML}{ff6b6b}
\definecolor{ffd0d0}{HTML}{ffd0d0}
\definecolor{fff1f1}{HTML}{fff1f1}
\definecolor{ffa9a9}{HTML}{ffa9a9}
\definecolor{ffacac}{HTML}{ffacac}
\definecolor{ff6e6e}{HTML}{ff6e6e}
\definecolor{ffdbdb}{HTML}{ffdbdb}
\definecolor{ff6a6a}{HTML}{ff6a6a}
\definecolor{ff4949}{HTML}{ff4949}
\definecolor{fff2f2}{HTML}{fff2f2}
\definecolor{ffaeae}{HTML}{ffaeae}
\definecolor{ff7272}{HTML}{ff7272}
\definecolor{ffe3e3}{HTML}{ffe3e3}
\definecolor{fffafa}{HTML}{fffafa}
\definecolor{ffcdcd}{HTML}{ffcdcd}
\definecolor{ffcaca}{HTML}{ffcaca}
\definecolor{fff7f7}{HTML}{fff7f7}
\definecolor{fff0f0}{HTML}{fff0f0}
\definecolor{ffdcdc}{HTML}{ffdcdc}
\definecolor{ff2d2d}{HTML}{ff2d2d}
\definecolor{ff6363}{HTML}{ff6363}
\definecolor{ff8686}{HTML}{ff8686}
\definecolor{ff0202}{HTML}{ff0202}
\definecolor{ffeded}{HTML}{ffeded}
\definecolor{ff0303}{HTML}{ff0303}
\definecolor{fff3f3}{HTML}{fff3f3}
\definecolor{ff0e0e}{HTML}{ff0e0e}
\definecolor{fff6f6}{HTML}{fff6f6}
\definecolor{fff9f9}{HTML}{fff9f9}
\definecolor{ff9e9e}{HTML}{ff9e9e}
\definecolor{ff2b2b}{HTML}{ff2b2b}
\definecolor{ffcece}{HTML}{ffcece}
\definecolor{ff2a2a}{HTML}{ff2a2a}
\definecolor{ff2525}{HTML}{ff2525}
\definecolor{fff5f5}{HTML}{fff5f5}
\definecolor{ff6262}{HTML}{ff6262}
\definecolor{ffb7b7}{HTML}{ffb7b7}
\definecolor{ffd8d8}{HTML}{ffd8d8}
\definecolor{ffbcbc}{HTML}{ffbcbc}
\definecolor{ff4e4e}{HTML}{ff4e4e}
\definecolor{fffbfb}{HTML}{fffbfb}
\definecolor{ffe0e0}{HTML}{ffe0e0}
\definecolor{ffd4d4}{HTML}{ffd4d4}
\definecolor{ff1414}{HTML}{ff1414}
\definecolor{ff1616}{HTML}{ff1616}
\definecolor{ffd3d3}{HTML}{ffd3d3}
\definecolor{ff8c8c}{HTML}{ff8c8c}
\definecolor{ffb3b3}{HTML}{ffb3b3}
\definecolor{ff6c6c}{HTML}{ff6c6c}
\definecolor{ffefef}{HTML}{ffefef}
\definecolor{ff7070}{HTML}{ff7070}
\definecolor{e5e5e5}{HTML}{e5e5e5}
\definecolor{cccccc}{HTML}{cccccc}
\definecolor{b2b2b2}{HTML}{b2b2b2}

\PassOptionsToPackage{hyphens}{url}
\usepackage[colorlinks = true,
            linkcolor = blue,
            urlcolor  = blue,
            citecolor = blue,
            anchorcolor = blue]{hyperref}
\usepackage{etoolbox}

\renewenvironment{abstract}
  {{\bfseries\noindent{\abstractname}\par\nobreak}\footnotesize}
  {\bigskip}

  {\begin{tabular}{|p{13cm}}}
  {\end{tabular}}
  
\titlespacing{\section}{0pt}{*3}{*1}
\titlespacing{\subsection}{0pt}{*2}{*0.5}
\titlespacing{\subsubsection}{0pt}{*1.5}{0pt}

\usepackage{authblk}

\providecommand\citet{\cite}
\providecommand\citep{\cite}

\newif\iflatexml\latexmlfalse

\begin{document}

\title{Multilayer Horizontal Visibility Graphs for Multivariate Time Series Analysis\footnote{Accepted authors manuscript (AAM) published in Data Mining and Knowledge Discovery (2025) 39:17. Final source webpage: \url{https://doi.org/10.1007/s10618-025-01089-4}. CONTACT Vanessa Freitas Silva. Email: \url{vanessa.silva@fc.up.pt} }}

\author[1]{Vanessa Freitas Silva}%
\author[2]{Maria Eduarda Silva}%
\author[1]{Pedro Ribeiro}%
\author[1]{Fernando Silva}%

\affil[1]{CRACS-INESC TEC, Faculdade de Ciências, Universidade do Porto}%
\affil[2]{LIAAD-INESC TEC, Faculdade de Economia, Universidade do Porto}%

\vspace{-1em}

\date{}

\begingroup
\let\center\flushleft
\let\endcenter\endflushleft
\maketitle
\endgroup

\selectlanguage{english}
\begin{abstract}
Multivariate time series analysis is a vital but challenging task, with multidisciplinary applicability, tackling the characterization of multiple interconnected variables over time and their dependencies. Traditional methodologies often adapt univariate approaches or rely on assumptions specific to certain domains or problems, presenting limitations. A recent promising alternative is to map multivariate time series into high-level network structures such as multiplex networks, with past work relying on connecting successive time series components with interconnections between contemporary timestamps.

In this work, we first define a novel cross-horizontal visibility mapping between lagged timestamps of different time series and then introduce the concept of multilayer horizontal visibility graphs. This allows describing cross-dimension dependencies via inter-layer edges, leveraging the entire structure of multilayer networks. To this end, a novel parameter-free topological measure is proposed and common measures are extended for the multilayer setting. Our approach is general and applicable to any kind of multivariate time series data.

We provide an extensive experimental evaluation with both synthetic and real-world datasets. We first explore the proposed methodology and the data properties highlighted by each measure, showing that inter-layer edges based on cross-horizontal visibility preserve more information than previous mappings, while also complementing the information captured by commonly used intra-layer edges. We then illustrate the applicability and validity of our approach in multivariate time series mining tasks, showcasing its potential for enhanced data analysis and insights.

\end{abstract}%

\par\null

\section{Introduction}\label{sec1}

Recent advancements in technology have led to the widespread availability of substantial amounts of high-dimensional time-indexed data. To effectively analyze such multidimensional time-indexed data, commonly referred to as multivariate time series, appropriate methodological and computational tools are essential. These data types are prevalent across diverse domains, including climate studies, health monitoring, and financial data analysis. Multivariate time series are characterized by serial correlation and cross-sectional dependencies, often termed the  "curse of dimensionality"~\citep{wei2018multivariate}.

\newpage
To address the challenge of analyzing high-dimensional time-indexed data, feature-based approaches have been proposed. Traditionally, time series features rely on statistics and models that may involve pre-processing and assumptions not universally satisfied. An alternative methodology involves mapping time series into complex networks~\citep{vanessa2020} to extract topological features for mining tasks and forecasting~\citep{zhang2018forecasting,li2021multivariate,vanessa2022}. Network science, which studies information extraction from complex networks, offers a rich set of topological graph measurements~\citep{Albert2002,Costa2007,peach2021hcga}.

Univariate time series can typically be mapped into single-layer networks utilizing concepts such as visibility, transition probability, and proximity~\citep{zou2018complex,vanessa2020}. On the other hand, when dealing with multivariate time series, one may opt for mapping these into either single-layer or multiple-layer networks. In the former, nodes symbolize the individual time series components, and the edges represent relationships between the latter, computed using statistical methods or models. However, this approach often results in the loss of vital information regarding the dynamics of each time series component, including serial correlation.

Recognizing this limitation, several authors proposed mappings that represent multivariate time series as multiplex networks, aiming to preserve both the temporal dynamics and cross-sectional information within the data~\citep{lacasa2015network,sannino2017visibility,eroglu2018multiplex,vanessa2020}. In these multiplex networks, each univariate time series component is allocated to a distinct layer, employing a univariate time series mapping where each node represents a timestamp or pattern. Layers are interconnected through common nodes in successive layers. Nevertheless, this mapping process invariably leads to the loss of directed lagged cross-correlations, which are of considerable importance in most multivariate scenarios.

To overcome the limitations mentioned above, we propose a new mapping method to represent multivariate time series as multilayer complex networks. Multilayer networks, with their capacity for internal connections within the same layer and external connections between different layers, offer a comprehensive and flexible data representation  structure~\citep{kivela2014multilayer}. The new mapping relies on a new horizontal visibility concept, \textit{Cross-Horizontal Visibility}, designed to capture cross-dependencies between pairs of time series components. Extending previous work on multiplex visibility graphs~\citep{lacasa2015network,sannino2017visibility}, we incorporate \textit{inter-layer edges} based on cross-horizontal visibility. These connections capture dependencies between different (lagged) timestamps of various variables, resulting in what we term \textit{Multilayer Horizontal Visibility Graphs}.

In addition to accurate analysis and understanding of data, temporal data statistics play a crucial role in many time series analysis applications~\citep{fulcher2014highly,hyndman2015large,kang2020gratis}, particularly for dimensionality reduction.  Identifying a collection of features that encapsulates the main properties of such data is a pivotal task, which typically entails conventional statistical methods and non-linear measures~\citep{montero2020fforma}. While univariate time series data offers a broad range of descriptive features, including linear statistics, nonlinear dynamics, symbolization, and innovative topological features, features for multivariate time series are currently more limited and pose a challenge. To address this gap, we propose a set of global topological multilayer network features as a novel set for multivariate time series analysis. These features include \textit{intra-layer}, \textit{inter-layer}, \textit{all-layer} topological features, along with \textit{relational} features. Within the category of relational features, which pertains to topological features related to components of the network, we introduce a new feature designed to capture the relationship between intra-layer and inter-layer connections.

\begin{figure}[!ht]
\centering
\includegraphics[width=1\textwidth]{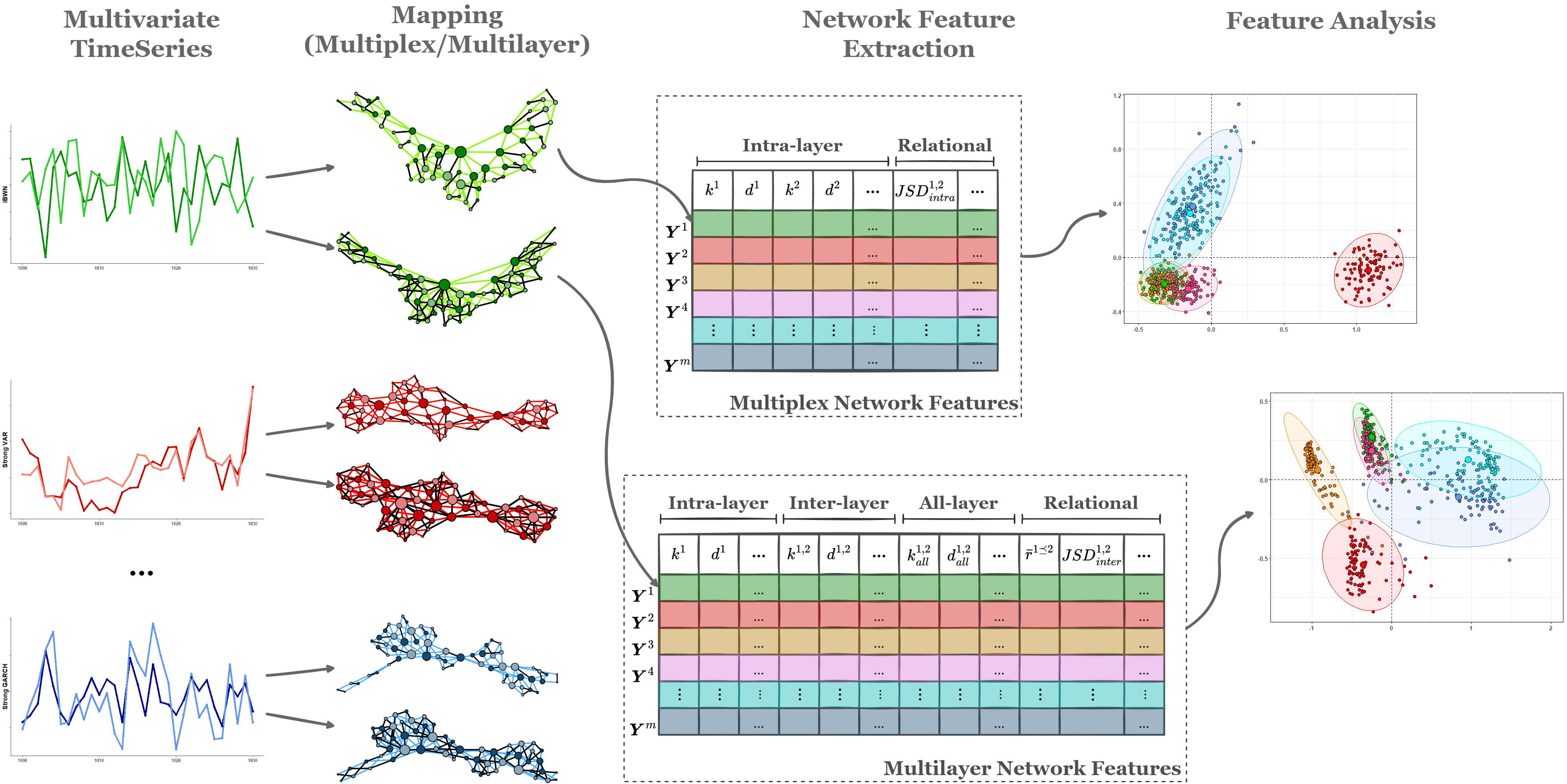}
\caption{Schematic diagram of the network-based features approach to time series mining tasks. The first column displays a set of synthetic multivariate time series, the second column the corresponding multiplex horizontal visibility graphs (based on~\citealp{lacasa2015network}) and the multilayer horizontal visibility graphs proposed in this work, the third column shows the set of topological features extracted from the two types of multiple-layer networks, and the last column a principal component analysis derived from the extracted topological feature sets. The description of the notation corresponding to the topological features presented in the third column is given in Section~\ref{sec4}.} 
\label{fig:overview_int}
\end{figure}

\newpage
The methodology proposed in this work and represented in Figure~\ref{fig:overview_int}, introduces a novel algorithmic method to transform multivariate time series data into multilayer networks, enabling the extraction of features from the resulting networks for comprehensive multivariate time series data analysis. This non-parametric methodology does not require data preprocessing and is not tailored to a specific time series mining problem or dataset, ensuring its versatility across various applications.
To comprehensively analyze, assess, and test the framework introduced in this work, we leverage synthetic multivariate time series generated from a selected set of multivariate time series models. Additionally, a diverse set of real-world multivariate time series, with varying characteristics and dimensionality properties, is employed in our evaluation.

\subsection{Contributions}
The main contributions of this work are the following:

\begin{itemize}
    \item \textbf{The Cross-Horizontal Visibility concept and mapping}. We propose a novel visibility concept that extends horizontal visibility defined within a time series to a bivariate setting. Based on cross-horizontal visibility we define a mapping for multivariate time series into multilayer visibility networks, by introducing inter-layer edges between different nodes of distinct layers. Multiplex visibility networks, then, become a particular case of the multilayer visibility networks. This novelty, to the best of our knowledge, has not been previously explored in the literature for the analysis of multivariate time series data~\footnote{Initially introduced in~\cite{MHVG2MTS_Arxiv23}}.
    
    \item \textbf{A new set of multidimensional features}. We introduce a new topological feature tailored for multilayer networks and curate a distinct set of multilayer network topological features for analyzing the proposed mapping method and reducing the dimensionality of multivariate time series data. We present a diverse set of topological features of multilayer networks, leveraging their intra- and inter-layer connections within the context of multivariate time series analysis. 
    
    \item \textbf{A comprehensive exploratory and empirical analysis}. We perform a thorough exploratory and empirical analysis of multidimensional topological features using synthetic datasets that encompass multiple correlation scenarios (both serial and crossed).    
    
    \item \textbf{Application of multidimensional time series features to mining tasks}. While the proposed multivariate time series mapping and the underlying features are not explicitly tailored for specific mining tasks, we demonstrate their application through clustering and classification tasks on a diverse set of synthetic and real-world datasets. It is important to note that this work primarily aims to introduce our proposed method, highlighting its versatility and ease of application. The emphasis is on presenting its practical utility rather than engaging in direct competition or comparison with other potential approaches.
    
\end{itemize}

\subsection{Organization}
We have structured this document as follows. 
Section~\ref{sec2} introduces fundamental concepts of multivariate time series and multilayer networks, establishing the notation for the paper. This section also provides background information on mapping methods crucial for comprehending the proposed methodology. 
Section~\ref{sec3} presents the concept of cross-horizontal visibility between time series components and introduces our novel multivariate time series mapping. 
In Section~\ref{sec4} we present the set of topological features extended to multilayer networks and, consequently, to multivariate time series data. This section also unveils a new topological feature designed specifically for multilayer networks. 
Section~\ref{sec5} conducts a comprehensive study of the proposed mapping method through the analysis of the corresponding feature set, providing insights into the properties of the multivariate time series. Additionally, this section explores multivariate time series mining tasks across synthetic and real-world datasets of various dimensions and types. 
Finally, Section~\ref{sec6} presents the conclusions drawn from our research, along with additional considerations, and outlines potential avenues for future work and exploration.

\section{Background}\label{sec2}
This section introduces fundamental concepts on multivariate time series and multilayer networks, thus establishing the notation that will be used in the paper. 

\subsection{Multivariate Time Series}
\label{subsec:mts}

An \textit{Univariate Time Series} (UTS) is a sequence of (scalar) observations indexed by time $t$, typically denoted as $\{{Y}_t\}_{t=1}^{T}$. Unlike a random sample, such observations are ordered in time and usually exhibit serial correlation that must be accounted for in the analysis. In instances where, at each time $t$, a vector of $m$ observations is obtained, expressed as $\boldsymbol{Y}_t = [Y_{1,t}, Y_{2,t}, \ldots, Y_{m,t}]^{\prime}$, where $\prime$ represents the transpose, the resulting dataset $\boldsymbol{Y}=\{\boldsymbol{Y}_t\}_{t=1}^{T}$ is termed a \textit{Multivariate Time Series} (MTS). Henceforward, the UTS components of the MTS $\boldsymbol{Y}$ are denoted by $\boldsymbol{Y}^{\alpha}=[Y_{\alpha,1}, Y_{\alpha,2}, \ldots, Y_{\alpha, T}]$, $\alpha=1, \ldots,m$ and thus we can denote the MTS by its components,  $\boldsymbol{Y}=\{Y^{\alpha}\}_{\alpha=1}^{m}$. MTS data present not only serial correlation within each component, $\boldsymbol{Y}^{\alpha}$, but also a correlation between the different UTSs, $\boldsymbol{Y}^{\alpha}$ and $\boldsymbol{Y}^{\beta}$ with $\alpha \neq \beta$, both \textit{contemporaneous} and \textit{lagged} correlation. Thus, analyzing MTS depends on key dependence measures such as the \textit{autocorrelation function} (ACF), which measures the linear predictability of a UTS,  
\begin{equation}
	\rho(s,t) = \mbox{\rm corr}(Y_t, Y_s) =  \frac{\mbox{\rm cov}(Y_s, Y_t)}{\sqrt{\mbox{\rm var}(Y_s) \mbox{\rm var}(Y_t)}},
\end{equation}
and the \textit{cross-correlation function} (CCF), which measures the correlation between any two components of the MTS, $\alpha$ and $\beta$, say, at times $s$ and $t$,
\begin{equation}
	\rho_{\alpha,\beta}(s,t) = \mbox{\rm corr}(Y_{\alpha,s}, Y_{\beta,t}).
\end{equation}

Time series analysis encompasses a range of methodologies developed to systematically address statistical challenges posed by serial correlation. A multitude of statistical models, both linear and non-linear, are available in the literature for effectively characterizing the dynamics of UTS~\citep{Sumway2017}. While the theoretical framework for UTS naturally extends to the multivariate case, incorporating concepts such as mean, covariance, ACF, and CCF, new complexities emerge. MTS analysis demands specialized tools, methods, and models capable of extracting valuable insights from multiple measurements exhibiting both temporal and cross-sectional correlations. This requires the development of innovative approaches to effectively capture the intricacies inherent in multivariate time series data.

\subsection{Multilayer Networks}
\label{subsec:mnet}

A \textit{graph} (or \textit{network}) $G$, is a mathematical structure defined by a pair $(V, E)$, where $V$ represents the set of \textit{nodes} and $E$ the set of \textit{edges} (connections) between pairs of nodes. Two nodes $v_i$ and $v_j$ are called neighbors if they are connected, $(v_i,v_j) \in E$. If there is no direction from a source node to a target node the edges are undirected: $(v_i,v_j) \in E$ implies that $(v_j,v_i) \in E$. A graph can be represented by an \textit{adjacency matrix} $\boldsymbol{A}$, where $A_{i,j}$ is $1$ when $(v_i,v_j) \in E$ and is $0$ otherwise.

A \textit{Multilayer Network} (MNet) is a complete and general structure suitable for modeling multiple complex systems through their interactions, intra- and inter-systems. An MNet is generally defined as a quadruplet $M = (V_M, E_M, V, \boldsymbol{L})$ where $V$ and $\boldsymbol{L}$ represent the set of entities and the set of layers of $M$, respectively, and $V_M$ and $E_M$ represent the global sets of nodes and edges, respectively. The $V_M \subseteq V \times L_1 \times \ldots \times L_m$, where $L_{\alpha} \in \boldsymbol{L}$ is an elementary layer, is a set of node-layer combinations in which a node is present in the corresponding layer  $L_{\alpha}$. The $E_M \subseteq V_M \times V_M$ is the set of edges that contain the pairs of possible combinations of nodes and elementary layers~\citep{kivela2014multilayer}. We denominate as \textit{intra-layer edges}, the connections between nodes of the same layer, $(v_i^{\alpha}, v_j^{\alpha})$, and \textit{inter-layer edges} the connections between nodes of different layers, $(v_i^{\alpha}, v_j^{\beta})$ with $\alpha \neq \beta$. 

\begin{figure}[!ht]
\centering
\includegraphics[scale=0.43]{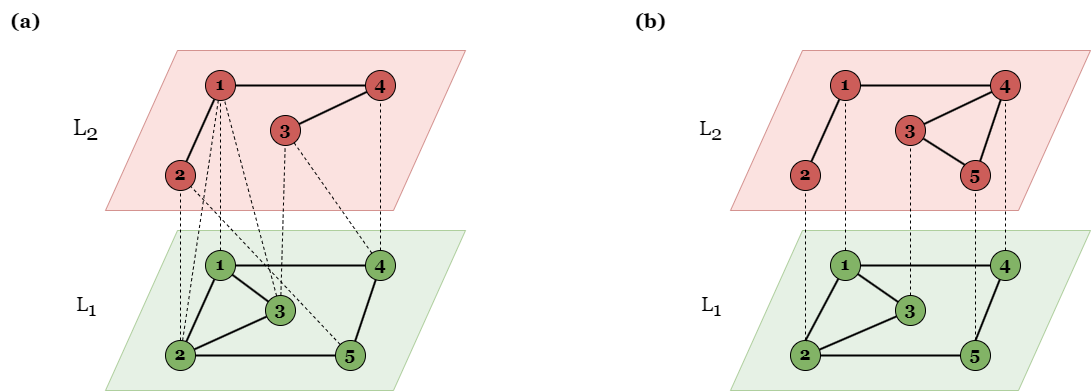}
\caption{Illustrative example of two toy multilayer networks with five entities, $V = \{1,2,3,4,5\}$, and two layers $\boldsymbol{L} = \{L_1, L_2\}$. (a) represents a toy multilayer network and (b) a toy multiplex network. Solid lines represent the intra-layer edges and dashed lines represent the inter-layer edges. \\ \textit{Source}: Modified from~\cite{vanessa2020}.} \label{fig:mnet}
\end{figure}

Two specific instances of multilayer networks include the \textit{monoplex network}, where $m=1$ and $M$ reduces to a single-layer network denoted as $G$ and the multiplex network, where $M$ is a sequence of $m$ graphs, $\{G^\alpha\}_{\alpha=1}^m = \{(V^\alpha, E^\alpha)\}_{\alpha=1}^m$. In the case of \textit{multiplex networks}, these graphs typically share a common set of nodes across all elementary layers, and inter-layer edges exclusively connect corresponding nodes across all successive layers, i.e., connecting $(v_i^{\alpha}, v_i^{\beta})$ where $\alpha \neq \beta$~\citep{boccaletti2014structure}. Figure~\ref{fig:mnet} illustrates the representation of both simple multilayer and multiplex networks.

\begin{figure}[!ht]
\centering
\includegraphics[scale=0.43]{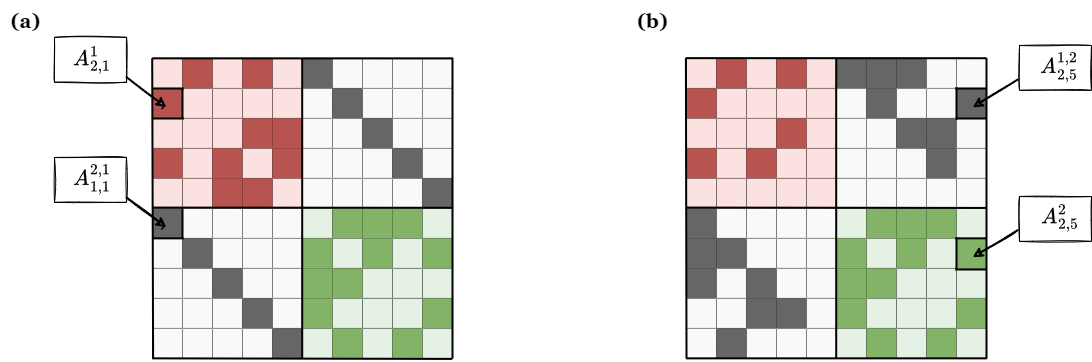}
\caption{Illustrative example featuring two supra-adjacency matrices: (a) depicting a supra-adjacency matrix of a toy multilayer network, and (b) showcasing a supra-adjacency matrix of a toy multiplex network. Colored blocks represent intra-layer graphs, while gray blocks represent inter-layer graphs.} \label{fig:adj_mnet}
\end{figure}

\newpage
A node-aligned MNet\footnote{A multilayer network is node-aligned if all layers contain all entities, that is, $V_M = V \times L_1 \times \ldots \times L_m$.} has an associated \textit{adjacency tensor} of order $4$, $\pmb{\mathcal{A}}$, with tensor elements $\mathcal{A}_{i,j,\alpha,\beta} = 1$ if $(v_i^{\alpha}, v_j^{\beta}) \in E_M$ and $0$ otherwise~\citep{kivela2014multilayer}. If the MNet is not node-aligned, we use empty nodes to fill the tensor structure. An alternative representation involves flattening $\pmb{\mathcal{A}}$ into a \textit{supra-adjacency matrix}, $\boldsymbol{A}$. In this matrix, intra-layer edges correspond to diagonal element blocks, while inter-layer edges are associated with off-diagonal element blocks. Figure~\ref{fig:adj_mnet} visually demonstrates the supra-adjacency matrices corresponding to the networks depicted in Figure~\ref{fig:mnet}. From these element blocks, we can deduce three types of subgraphs:
\begin{itemize}
    \item \textit{intra-layer graphs}, $G^{\alpha}$, represented by the square matrices of order $|V^{\alpha}|$ formed by the diagonal element blocks (intra-layer edges, $A_{i,j}^{\alpha}$), i.e., $\left[ \begin{smallmatrix}   \boldsymbol{A}^{\alpha} & \boldsymbol{0}\\ \boldsymbol{0}  & \boldsymbol{0} &  \end{smallmatrix} \right]$.
    
    \item \textit{inter-layer graphs}, $G^{\alpha,\beta}$, represented by the square matrices of order $|V^{\alpha}|+|V^{\beta}|$ constructed from off-diagonal element blocks (inter-layer edges, $A_{i,j}^{\alpha,\beta}$ and $A_{j,i}^{\beta,\alpha}$, and no intra-layer edges, $A_{i,j}^{\alpha} = 0$ and $A_{i,j}^{\beta} = 0$)~\footnote{Note also that the inter-layer graphs have the characteristics of a bipartite graph. A bipartite graph $G^{\alpha,\beta}$ has a node set $V^{\alpha,\beta}$ that can be divided into two disjoint and independent sets $V^{\alpha}$ and $V^{\beta}$ ($V^{\alpha,\beta} = V^{\alpha} \cup V^{\beta}$ and $V^{\alpha} \cap V^{\beta} = \emptyset$) and every edge connects a node in $V^{\alpha}$ to a node in $V^{\beta}$.}, i.e., $\left[ \begin{smallmatrix} \boldsymbol{0} & \boldsymbol{A}^{\alpha,\beta} \\ \boldsymbol{A}^{\beta,\alpha} & \boldsymbol{0}  \end{smallmatrix} \right]$. 
    
    \item \textit{all-layer graphs}, $G^{\alpha, \beta}_{all}$, represented by the square matrices of size $|V^{\alpha}|+|V^{\beta}|$ constructed by both on and off-diagonal element blocks (intra-layer edges, $A_{i,j}^{\alpha}$ and $A_{i,j}^{\beta}$, and inter-layer edges, $A_{i,j}^{\alpha,\beta}$ and $A_{j,i}^{\beta,\alpha}$), i.e., $\left[ \begin{smallmatrix} \boldsymbol{A}^{\alpha} & \boldsymbol{A}^{\alpha,\beta} \\ \boldsymbol{A}^{\beta,\alpha} & \boldsymbol{A}^{\beta}  \end{smallmatrix} \right]$. 
\end{itemize}

Network science stands as a valuable tool for addressing diverse problems in various scientific disciplines~\citep{vespignani2018twenty}. A plethora of topological, statistical, spectral, and combinatorial metrics, extracting information from networks, is available in the literature~\citep{Albert2002,Barabasi2016,Costa2007,peach2021hcga,vanessa2022}. These metrics can be categorized into global, local, and "intermediate" features. The first group quantifies properties involving all network elements, the second focuses on properties over a specific node or edge, and the last pertains to properties that involve subsets of the network, such as subgraphs. In multilayer contexts, these metrics find natural extensions to MNets. Most common topological metrics can be straightforwardly extended to 
\textit{intra-layer metrics} by computation over intra-layer edges. Furthermore, these can also be applied to the entire MNet by computation over both intra-layer and inter-layer edges~\citep{kivela2014multilayer,huang2021survey}. Other approaches rely on measurements and properties inspired in tensor analysis literature~\citep{kivela2014multilayer}.

\subsection{Mapping Time Series into Networks via Visibility Concepts}\label{subsec:mts_mnet}

Over the past two decades, numerous approaches for time series analysis grounded in network principles have emerged. These methodologies involve the transformation of both univariate and multivariate time series into the network domain, manifested as either single-layer or multiple-layer networks~\citep{vanessa2020}. 
The mappings presented in the literature are essentially based on concepts of visibility, transition probability, proximity, time series models, and statistics~\citep{zou2018complex,vanessa2020}. 
In this section, we review the concepts of horizontal visibility graphs and multiplex visibility graphs essential for the discussion that follows.

The \textit{Horizontal Visibility Graph} (HVG)~\citep{Luque2009} is a graph $G=(V,E)$ associated with a Univariate Time Series (UTS) $\{Y_t\}_{t=1}^T$, where $V=\{v_t, t=1,\ldots,T\}$ with $v_t$ representing the timestamp $t$. 
For all $i,j = 1 \ldots T, i \neq j, (v_i, v_j) \in E$ if, for all $k$ with $i < k < j$ the following condition on the UTS observations holds: 
\begin{equation}
Y_{k} < Y_{i} \quad \wedge  \quad Y_{k} <  Y_{j}. \label{eq_hvg}
\end{equation}
In particular, when $j = i+1, (v_{i}, v_{i+1}) \in E$ meaning that consecutive nodes are always connected.

\begin{figure}[!htbp]
\centering
\includegraphics[width=1\textwidth]{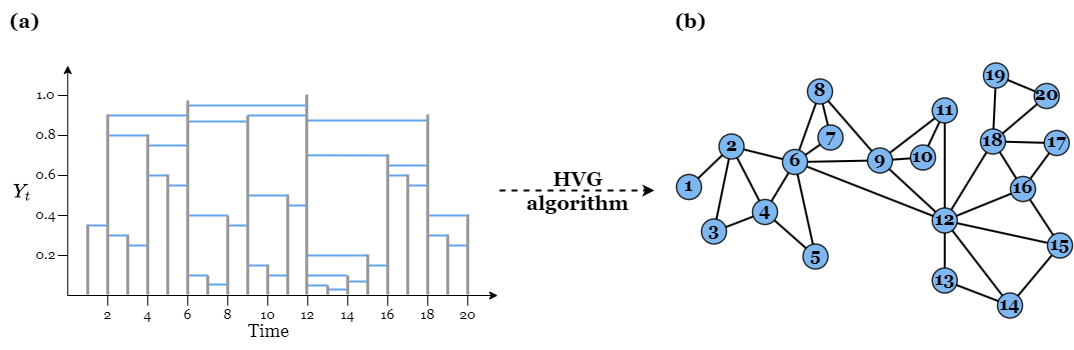}
\caption{Illustrative example of the Horizontal Visibility Graph (HVG) algorithm. (a) represents a toy time series and corresponding visibility between data bars (observations), where solid blue lines represent the horizontal visibility lines according to the HVG method. (b) represents a network generated by the corresponding mapping. \\ \textit{Source}: Adapted from~\cite{vanessa2020}.} \label{fig:hvg}
\end{figure}

Intuitively, the Horizontal Visibility Graph (HVG) algorithm 
  interprets each observation, denoted as $Y_t$, in a UTS as a vertical bar laid on a landscape, connecting different bars based on a visibility criterion. The height of each bar corresponds to its numerical value (at time $t$), and the visibility between two bars depends on their heights, determining mutual visibility. In an HVG, each timestamp, $t$, maps to a node, $v_t$, and edges $(v_i, v_j)$ for $i,j = 1 \ldots T, i \neq j$ are established if there is a horizontal visibility line between the corresponding data bars that is not intercepted by another data bar between the corresponding timestamps. 
Figure~\ref{fig:hvg} illustrates the HVG method.

\newpage

\paragraph{Some properties of HVG~\citep{Luque2009}:}
\vspace{-.5cm}
\begin{itemize}
    \item{\textit{Connected:}} HVGs are always connected graphs, as each node $v_i$ has horizontal visibility to its nearest neighbors $v_{i-1}$ and $v_{i+1}$. 
    \item{\textit{Undirected/Directed:}} HVGs edges are undirected unless we consider the direction of the time axis or another directional concept.
    \item{\textit{Invariant under affine transformations:}} Each transformation $X_t= aY_t + b$, where $a \in \mathbb{R}, b \in \mathbb{R}$, and $t = 1, \ldots, T$, leads to the same HVG.
\end{itemize}

The horizontal visibility method represents global and local topological characteristics of time series data in the graph. It is easy to implement, computationally fast, and parameter-free. In particular, there exists a set of mathematically well-defined relationships~\citep{Luque2009} between the properties of HVGs and the underlying time series characteristics (see~\citealp{vanessa2020} for more details).

\begin{figure}[!ht]
\centering
\includegraphics[width=1\textwidth]{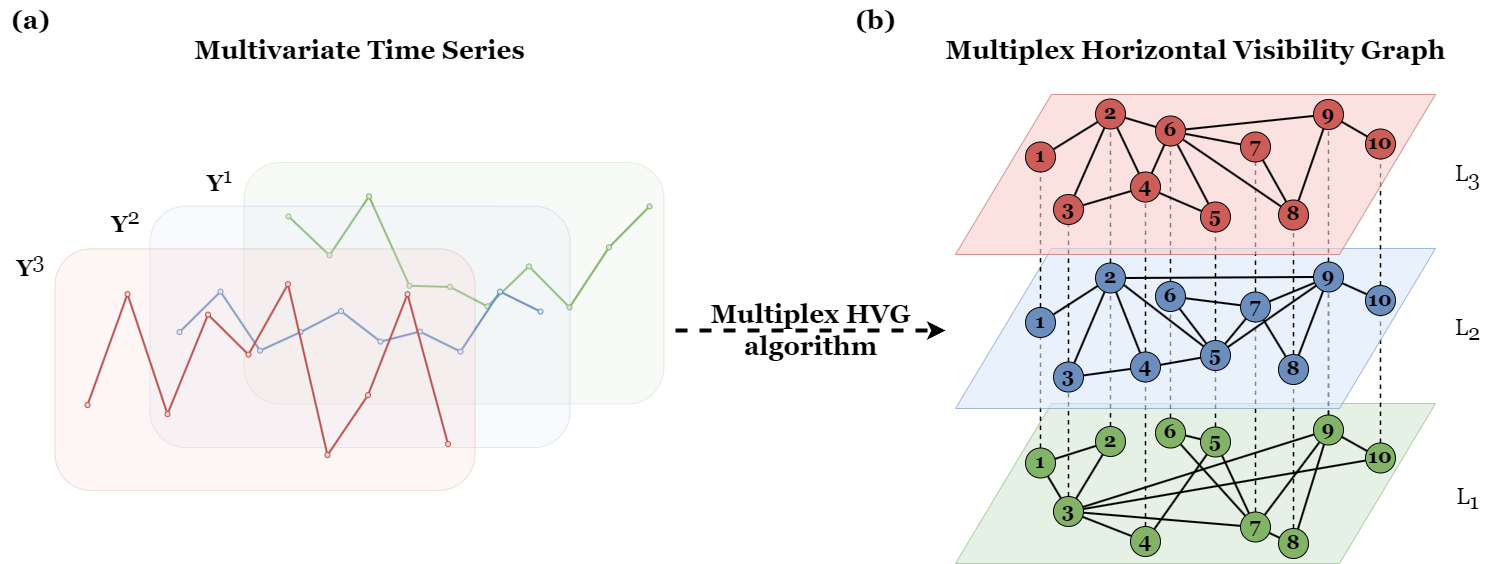}
\caption{Illustrative example of the Multiplex Horizontal Visibility Graph (HVG) algorithm: (a) represents a toy multivariate time series with three components, $\boldsymbol{Y} = \{\boldsymbol{Y}^{1}, \boldsymbol{Y}^{2}, \boldsymbol{Y}^{3}\}$, and (b) the resulting multiplex network with three layers generated by the Multiplex HVG mapping.\\ \textit{Source}: Adapted from~\cite{vanessa2020}.} \label{fig:vg_mplx}
\end{figure}

The \textit{Multiplex Visibility Graph} (MVG) algorithm was proposed by~\cite{lacasa2015network} as an extension of the horizontal visibility mapping for MTS analysis. 
Building upon the definition of MNet given in the previous section, formally, a MVG of $m$ layers, $M$, is constructed, where the layer set $\{L_{\alpha}\}_{\alpha = 1}^m$ corresponds to the HVGs, $\{G^\alpha\}_{\alpha = 1}^m$, associated with the time series components $\{Y^{\alpha}\}_{\alpha=1}^{m}$. The adjacency matrix vector $\boldsymbol{A}_M$, represents $M$ with elements being the adjacency matrices of each layer, $\boldsymbol{A}_M = \{\boldsymbol{A}^{1}, \ldots, \boldsymbol{A}^{m}\}$ where $A^{\alpha}_{i,j} = 1$ indicates that nodes $v_i^{\alpha}$ and $v_j^{\alpha}$ are connected in layer $L_{\alpha}$ and 0 otherwise. 
Figure~\ref{fig:vg_mplx} illustrates this method.

\section{\textit{MHVG:} a New Multivariate Time Series Mapping}\label{sec3}

Visibility mapping methods have demonstrated great promise in capturing time series properties that encompass both local and global characteristics of the data~\citep{vanessa2020}. 
For example, in~\cite{Lacasa2008}, the authors employ Natural Visibility Graph (NVG)\footnote{NVG differs from HVG in the visibility condition between two nodes (timestamps). While HVG uses horizontal visibility lines between corresponding data bars (Eq.~\ref{eq_hvg}), NVG employs a criterion based on direct visibility lines between the tops of the data bars~\citep{Lacasa2008}.} and demonstrate that periodic time series are mapped onto regular graphs, random time series are mapped onto random graphs, and fractal time series onto scale-free graphs. Notably, HVGs predominantly preserve local data information by connecting the closest timestamps due to their horizontal line visibility criterion. In our previous works~\citep{vanessa2018time,vanessa2022}, we emphasized crucial characteristics of time series data reflected in the topological features of HVGs. For instance, we correlated clustering-based measures with autocorrelation values in autoregressive data and with properties of discrete time series. Additionally, we observed that time series based on states and periodic data are mapped into HVGs exhibiting a topological structure that emphasizes communities, as reflected in relevant measures such as communities and centrality measures.

Approaches based on networks, particularly leveraging their topological features, have proven not only relevant but also complementary to conventional time series analysis. Various mining tasks for UTS and, more recently, for MTS have emerged, yielding promising results~\citep{fulcher2017hctsa,li2021multivariate,vanessa2022}. 
Recent literature indicates that multiplex versions of visibility graphs when applied to MTS data, can achieve enhanced accuracy and more promising outcomes compared to traditional time series mappings~\citep{lacasa2015network,sannino2017visibility,eroglu2018multiplex,vanessa2020}.

\newpage
However, to our knowledge, the aforementioned results do not directly incorporate inter-layer edges. The analysis involves examining the networks through topological measurements on the monoplex networks resulting from flattening approaches or using similarity measures on the individual layers of multiplex networks. We argue that high-level network structures can retain a greater amount of data information post-mapping functions. This structure also allows us to expand the range of resources available in the literature and explore additional network components, such as inter-layer edges, thereby broadening the exploration of multidimensional data.

This section introduces a novel visibility algorithm to map an MTS into a \textit{Multilayer Horizontal Visibility Graph} (MHVG). The algorithm is based on a new visibility concept called \textit{cross-horizontal visibility}, extending the traditional horizontal visibility.

\subsection{Cross-Horizontal Visibility}
\label{subsec:cross_hv}

\begin{figure}[!ht]
\centering
\includegraphics[width=1\textwidth]{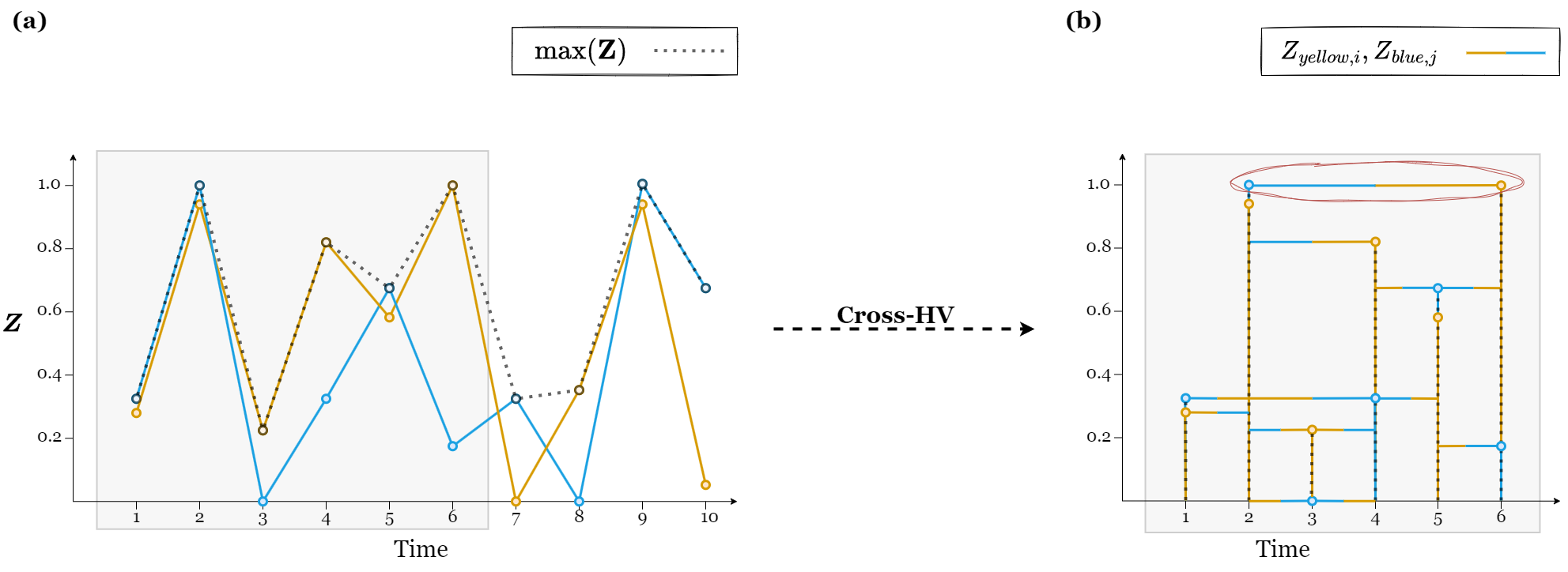}
\caption{Schematic diagram of the Cross-Horizontal Visibility (Cross-HV) concept. (a) Illustrates a toy bivariate time series $\boldsymbol{Z}^{yellow},\boldsymbol{Z}^{blue}$, (in the same scale), and the corresponding \textit{maximum} time series; (b) represents the cross-horizontal visibility by solid bi-color lines (yellow and blue) connecting the data bars of the time series components $\boldsymbol{Z}^{yellow}$ and $\boldsymbol{Z}^{blue}$, for the first six timestamps.} \label{cross_hv}
\end{figure}

Consider two time series $\boldsymbol{Z}^{\alpha}=\left(Z_{\alpha,1}, \ldots, Z_{\alpha,T}\right)$ and $\boldsymbol{Z}^{\beta}=(Z_{\beta,1}, \ldots, Z_{\beta,T})$ on the same scale. Two arbitrary data values 
$Z_{\alpha,i}$ 
and 
$Z_{\beta,j}$ 
are said to have cross-horizontal visibility, Cross-HV, if for all timestamps 
$k$ 
with 
$i < k < j$
and $i, j = 1, \ldots, T, i \neq j$, the following condition is satisfied:
\begin{equation}
    \max{\left(Z_{\alpha,{k}}, Z_{\beta,{k}}\right ) } < Z_{\alpha,i} \quad \wedge  \quad \max{\left(Z_{\alpha,{k}}, Z_{\beta,{k}}\right ) } < Z_{\beta,j}. 
    \label{eq_chvg}
\end{equation}
In the particular case where $j = i+1$ the corresponding data values  $Z_{\alpha,i}$ and $Z_{\beta,{i+1}}$ also have Cross-HV. 
Figure~\ref{cross_hv} illustrates the concept of Cross-HV with two toy time series.

Cross-HV condition establishes a direct geometric relationship between lagged timestamps of two variables using cross-horizontal visibility lines. The objective is to capture cross-lagged dependencies and joint dynamic properties between pairs of time series. Timestamps from two layers are then connected if the internal dynamics of both time series (i.e., between the specified timestamps) do not change in a way that restricts the cross-horizontal visibility criterion.

\newpage
\paragraph{Some properties of Cross-HV:}
\vspace{-.5cm}
\begin{itemize}
    \item{\textit{Reciprocal~\footnote{We use the term \textit{reciprocal} where \textit{mutual} is used in the literature since we find the former is a more appropriate term to describe the property as \textit{it implies a back-and-forth nature}.}:}} 
    It means that for all $\alpha,\beta,i,j$ with $\alpha \neq \beta, i \neq j$, if $Z_{\alpha,i}$ has Cross-HV to $Z_{\beta,j}$ then $Z_{\beta,j}$ has Cross-HV to $Z_{\alpha,i}$. In Figure~\ref{cross_hv} the first six data points are highlighted and bi-colored lines are used to indicate the (reciprocal) visibility. 
    \item{\textit{Non-Symmetric:}} Cross-HV is not symmetric. This means that for all $\alpha,\beta,i,j$ with $\alpha \neq \beta, i \neq j$, if $Z_{\alpha,i}$ has Cross-HV to $Z_{\beta,j}$, $Z_{\beta,i}$ may not have Cross-HV to $Z_{\alpha,j}$. Figure~\ref{cross_hv} illustrates the lack of symmetry in Cross-HV: the observation $Z_{blue,2}$ has Cross-HV to $Z_{yellow,6}$ (highlighted by the red ellipse), but 
    $Z_{yellow,2}$ does not have Cross-HV to $Z_{blue,6}$.
\end{itemize}

\subsection{Multilayer Horizontal Visibility Graph}
\label{subsec:mhvg}

\begin{figure}[!ht]
\centering
\includegraphics[width=1\textwidth]{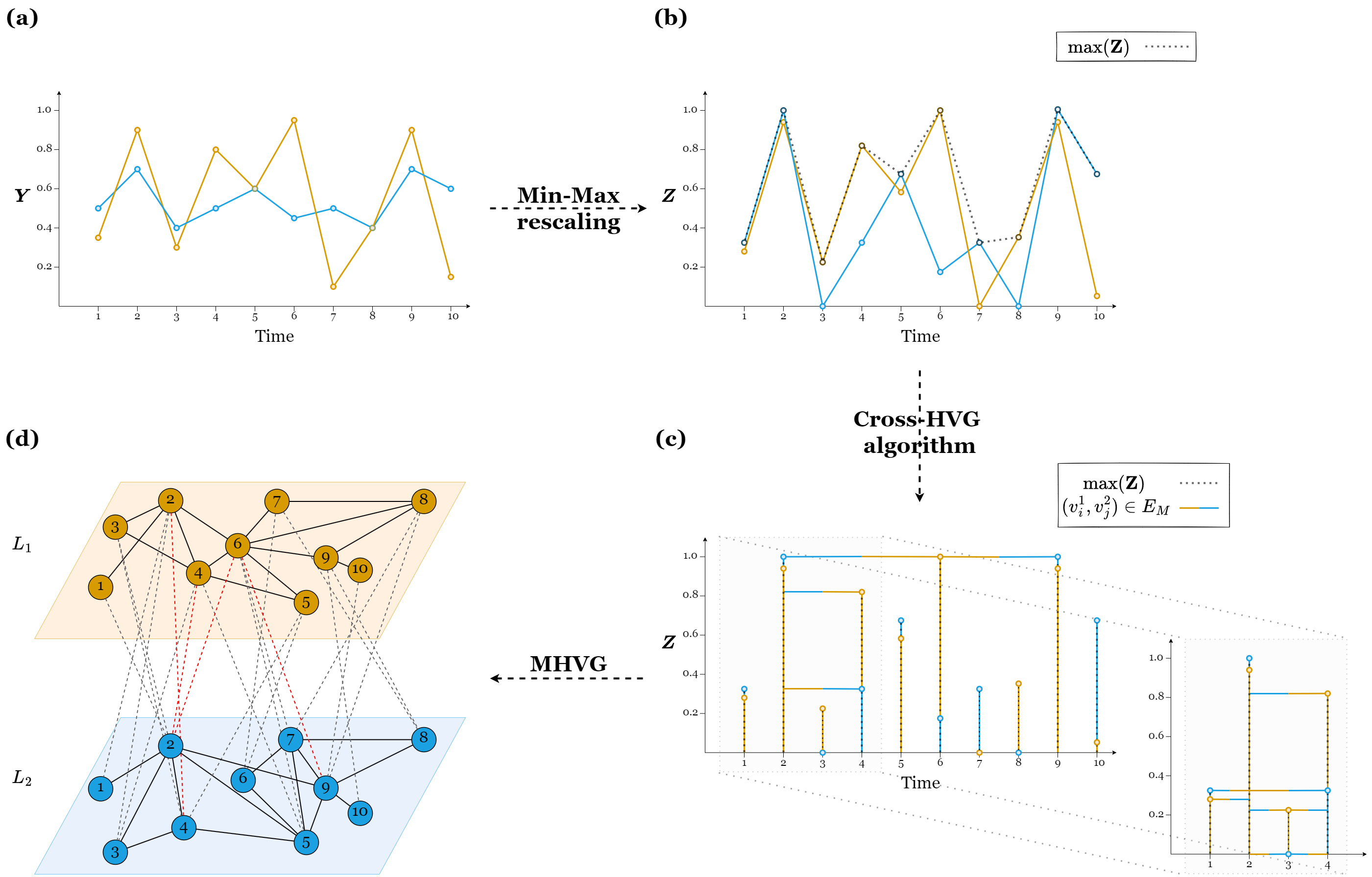}
\caption{Schematic diagram of the Multilayer Horizontal Visibility Graph (MHVG) algorithm: (a) represents the original toy time series, (b) shows the Min-Max re-scaled time series and \textit{maximum} time series, (c) showcases the Cross-Horizontal Visibility (Cross-HV) with the edges between adjacent timestamps omitted for simplicity (detail for the first four timestamps), and (d) illustrates the MHVG: solid black lines represent the intra-layer edges (the HVGs), dashed lines the inter-layer edges (the Cross-HVGs), and the red lines highlight inter-layer edges between nodes corresponding to non-adjacent timestamps.} \label{map_vg2}
\end{figure}

A \textit{Multilayer Horizontal Visibility Graph} (MHVG) is obtained by mapping an MTS, $\boldsymbol{Y}=\{Y^{\alpha}\}_{\alpha=1}^{m}$, into an MNet structure, $M = (V_M, E_M, V, \boldsymbol{L})$, using the concepts of HV and Cross-HV, as follows. Each unique timestamp, $t$, is mapped into an unique entity in $V_M$ and each component time series, $\boldsymbol{Y}^{\alpha}$, is mapped  into a layer, $L_{\alpha} \in \boldsymbol{L}$, $\alpha=1, \ldots, m$, using the HVG method described in Section~\ref{subsec:mts_mnet}, thus establishing the intra-layer edges, $(v_i^{\alpha}, v_j^{\alpha}) \in E_M$, $i,j=1, \ldots, T$, $i \neq j$. Then inter-layer edges $(v_i^{\alpha}, v_j^{\beta}) \in E_M, $ between any two layers $L_{\alpha}$ and $L_{\beta}$, $\alpha,\beta = 1, \ldots, m$, $ \alpha \neq \beta$ and $i,j = 1, \dots, T$, $i \neq j$ are established using the Cross-HV  described above in Section~\ref{subsec:cross_hv}. Note that to establish Cross-HV all the time series $\boldsymbol{Y}^{\alpha}$,  $\alpha=1, \ldots,m$ must be in the same scale, which may require a pre-processing step of the dataset $\boldsymbol{Y}$, comprising the Min-Max scaling of each time series. The mapping is illustrated in Figure~\ref{map_vg2}, with toy bivariate time series, for the sake of simplicity.

\begin{algorithm}[!t]
    \caption{Cross-Horizontal Visibility Graph}\label{alg:chvg2}

    \hspace*{\algorithmicindent} \textbf{Input:} {two rescaled time series ($\boldsymbol{tsA}$, $\boldsymbol{tsB}$), associated maximum time series ($\boldsymbol{tsMax}$), two horizontal visibility graph layers ($\boldsymbol{layerA}$, $\boldsymbol{layerB}$), and multilayer network ($\boldsymbol{mnet}$)}
    \\
    \hspace*{\algorithmicindent} \textbf{Output:} {inter-layer edges between the two layers ($layerA$ and $layerB$) of $mnet$}

    \begin{algorithmic}[1]

    \Procedure{CHVG}{$tsA, tsB, tsMax, layerA, layerB, mnet$} 
        \State $T \gets tsMax.size()$ 
        \For{$i \gets 1$ \textbf{to} $T-1$} \label{alg1_line3}
            \For{$j \gets i+1$ \textbf{to} $T$}
                \If{$tsMax[k] < tsA[i]$ \textbf{and} $tsMax[k] < tsB[j]$ \textbf{for all} $i<k<j$}
                    \State $mnet.$\Call{add\_Edge}{$i, j, layerA, layerB$}
                \EndIf
            \EndFor
        \EndFor \label{alg1_line9}
        \For{$i \gets 2$ \textbf{to} $T$} \label{alg1_line10}
            \For{$j \gets i-1$ \textbf{to} $1$}
                \If{$tsMax[k] < tsA[i]$ \textbf{and} $tsMax[k] < tsB[j]$ \textbf{for all} $i<k<j$}
                    \State $mnet.$\Call{add\_Edge}{$i, j, layerA, layerB$}
                \EndIf
            \EndFor
        \EndFor \label{alg1_line16}

        \State \Return{}
    \EndProcedure
    \end{algorithmic}
\end{algorithm}

From the generated MHVG, we can identify the \textit{intra-layer graphs}, $\{G^{\alpha}\}_{\alpha = 1}^m$ and the \textit{inter-layer graphs,} $G^{\alpha, \beta}$, for $\alpha, \beta = 1, \ldots, m$ and $\alpha \neq \beta$. The former corresponds to the HVGs associated with each time series component and it is represented by the adjacency matrix $\boldsymbol{A}^{\alpha}$ with $A_{i,j}^{\alpha} = 1$ if  $(v_i^{\alpha},v_j^{\alpha}) \in E_M$ and 0 otherwise. The second set of graphs corresponds to the Cross-Horizontal Visibility Graphs (Cross-HVG) associated with each pair of different time series components and it is represented by the adjacency matrix $\boldsymbol{B}^{\alpha,\beta} = \left[ \begin{smallmatrix} \boldsymbol{0} & \boldsymbol{A}^{\alpha,\beta} \\ \boldsymbol{A}^{\beta,\alpha} & \boldsymbol{0} \end{smallmatrix} \right]$ with $A_{i,j}^{\alpha, \beta} = 1$ and $A_{j,i}^{\beta, \alpha} = 1$ if $(v_i^{\alpha},v_j^{\beta}) \in E_M$ and 0 otherwise. With this representation, we aim to map the serial dependencies within the data onto intra-layer graphs and the cross-dependencies onto inter-layer graphs. Consider the hypothetical case (trivial, but not real) of a two-component MTS where both components precisely represent the same values, that is, exhibiting the same behavior and correlations. In such an instance, the intra-layer adjacency matrices ($\boldsymbol{A}^{1}$ and $\boldsymbol{A}^{2}$), as well as the inter-layer adjacency matrices ($\boldsymbol{A}^{1,2}$ and $\boldsymbol{A}^{2,1}$) resulting from the MHVG mapping, would be identical, reflecting exactly the same correlations.

Algorithm~\ref{alg:chvg2} describes the Cross-HV method which establishes the inter-layer edges between two given time series. This involves two for-loops that check for Cross-HV between a pair of time series components  (Eq.~\ref{eq_chvg}). The first for-loop (lines~\ref{alg1_line3} to~\ref{alg1_line9}) tests the Cross-HV from a given node $v_i^{\alpha}$ to the nodes $v_j^{\beta}$ (with $j>i$) to its right (i.e., for timestamps newer than it), while the second for-loop (lines~\ref{alg1_line10} to~\ref{alg1_line16}) tests the Cross-HV from a given node $v_i^{\alpha}$ to the nodes $v_j^{\beta}$ (with $j<i$) to its left (i.e., for timestamps older than it). For the sake of simplicity, Algorithm~\ref{alg:chvg2} does not include all the implementation details related to the Cross-HV condition. In the Appendix~\ref{app:mhvg} Algorithm~\ref{alg:chvg}, we provide the same algorithm with the necessary details for reproducibility. 

\begin{algorithm}[!t]
    \caption{Multilayer Horizontal Visibility Graph}\label{alg:mhvg}

    \hspace*{\algorithmicindent} \textbf{Input:} {multivariate time series ($\boldsymbol{mts}$)}
    \\
    \hspace*{\algorithmicindent} \textbf{Output:} {multilayer network ($\boldsymbol{mnet}$)}
    \\
    \hspace*{\algorithmicindent} {\textsc{\textbf{HVG}} procedure (proposed in~\citealp{Lan2015}) creates a HVG from a time series}

    \begin{algorithmic}[1]

    \Procedure{MHVG}{$mts$} 
        \State $m \gets mts.size()$ 
        \State $mnet \gets \{\}$ 
        \State $n\_mts \gets \{\}$ 
        
        \For{$a \gets 1$ \textbf{to} $m$} \label{alg2_line5}
            \State $mnet.layers[a] \gets \{\}$ 
            \State \Call{HVG}{$mts[a], mnet.layers[a], mnet$} \label{alg2_line7} 
            \State $n\_mts[a] \gets$ \Call{MinMax}{$mts[a]$} \label{alg2_line8} 
        \EndFor \label{alg2_line9}

        \For{$a \gets 1$ \textbf{to} $m-1$} \label{alg2_line10}
            \State $layerA \gets mnet.layers[a]$
            \For{$b \gets a+1$ \textbf{to} $m$}
                \State $layerB \gets mnet.layers[b]$
                \State $tsMax \gets$ \Call{Max}{$n\_mts[a], n\_mts[b]$} \label{alg2_line14}
                \State \Call{CHVG}{$n\_mts[a], n\_mts[b], tsMax, layerA, layerB, mnet$} \label{alg2_line15} 
            \EndFor
        \EndFor \label{alg2_line17}

        \State \Return{$mnet$} \label{alg2_line18}

    \EndProcedure

    \end{algorithmic}
\end{algorithm}

Algorithm~\ref{alg:mhvg} describes the complete mapping of a multivariate time series into a Multilayer Horizontal Visibility Graph. The procedure starts by mapping each time series component of an MTS into the respective HVGs (following the condition of Eq.~\ref{eq_hvg}) and allocating them to individual layers on an MNet (line~\ref{alg2_line7}). Next, all series are re-scaled to the range $[0,1]$ using Mix-Max scaling (line~\ref{alg2_line8}). Finally, for pairwise time series components, the corresponding maximum time series is obtained (line~\ref{alg2_line14}), and the Cross-HV method is applied to establish the inter-layer connections (line~\ref{alg2_line15}), resulting in the creation of the MHVG corresponding to the given MTS (line~\ref{alg2_line18}).

The Cross-HV method requires time series to be on the same scale because the visibility method is sensitive to their values. If one series has much larger values, connections can only be made between the nearest neighbors of a given observation, as higher values can block visibility. We use the Min-Max method to scale the series to the same range. 
However, this scaling does not handle high noise levels well, and noisy data should be cleaned before scaling. If noise is low, the method maps the time series behavior accurately. Future work will explore the impact of noise on this method.

\subsection{MHVG: Computational Complexity}

The computational complexity of the proposed mapping method depends on two variables: $T$, the time series length, and $m$ the number of variables. It is determined by the two for-loops (lines~\ref{alg2_line5} to~\ref{alg2_line9} and lines~\ref{alg2_line10} to~\ref{alg2_line17} of Algorithm~\ref{alg:mhvg}) and the procedures performed within them. The computational complexity of the first for-loop is $\mathcal{O}(m(T\log T))$, as the loop iterates $m$ times (representing each time series component), and for each iteration, the HVG algorithm is executed. Using the version proposed in~\cite{Lan2015}, this incurs a complexity of $\mathcal{O}(T\log T)$. The second for-loop consists of two nested loops iterating through pairs of time series components. In each iteration, the call to the $\Call{CHVG}{}$ function determines the complexity of the loop. The proposed Cross-HVG mapping has a complexity $\mathcal{O}(T^2)$, which is determined by the for-loops that test the cross-horizontal visibility. Thus, the total complexity of the second for-loop in Algorithm~\ref{alg:mhvg} is $\mathcal{O}(m^2T^2)$. Finally, Algorithm~\ref{alg:mhvg} has computational complexity $\mathcal{O}(m^2T^2)$ which is determined by the procedure that tests Cross-HV between all pairs of time series in an MTS.

It is worth noting that the complexity of Algorithm~\ref{alg:chvg} can be improved by applying the divide-and-conquer strategy proposed by~\cite{Lan2015}. This improvement would result in a final complexity of the MHVG method of $\mathcal{O} (m^2 (T \log T))$. This improvement is particularly noteworthy in practical scenarios since, in real datasets, the variable $T$ is typically much larger than the variable $m$. Additionally, given the independence of the individual HVG procedures (for each layer) and Cross-HVG procedures (for each pair of layers), parallelization techniques can also improve the runtime complexity.

\begin{figure}[!ht]
\centering
\includegraphics[width=1\textwidth]{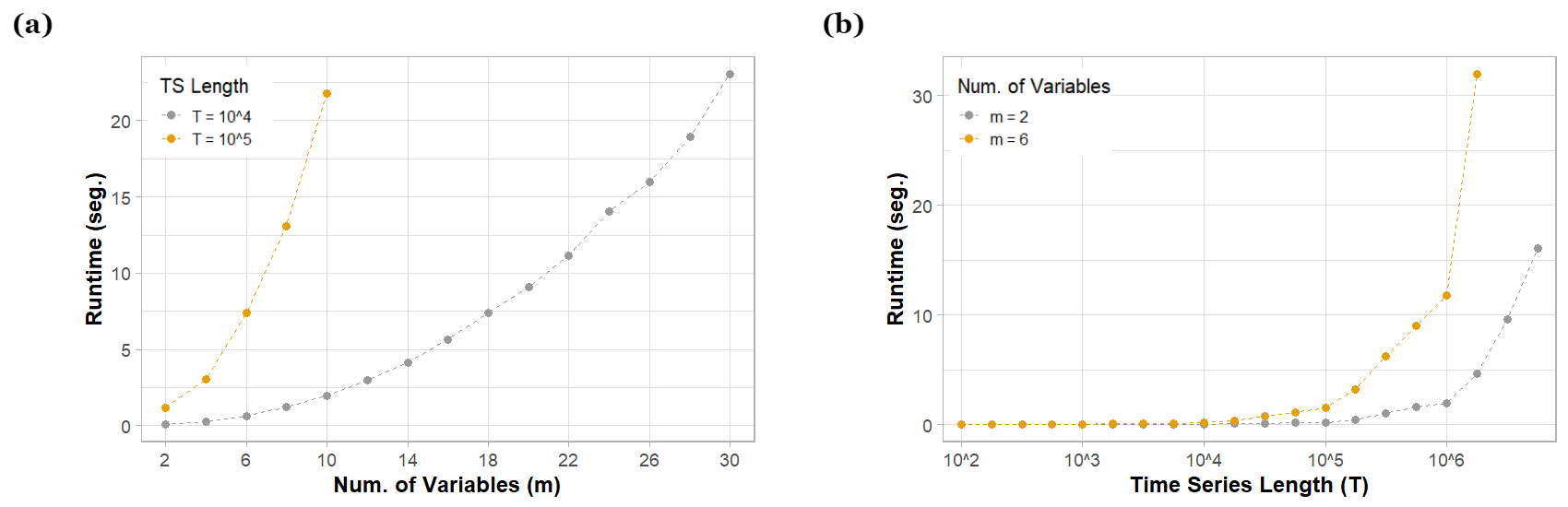}
\caption{Illustration of time complexity behavior for the MHVG mapping with synthetic data (independent white noises). Runtime, in seconds, for (a)  two fixed time series lengths, $T = 10^4$ and $T = 10^5$, and increasing the number of time series components $m$, ranging from $2$ to $30$; (b) fixed number of components, $m = 2$ and $m = 6$, and increasing time series length, $T$, from $10^2$ to $7.5\times10^6$.} \label{fig:runtimes}
\end{figure}

\newpage
Figure~\ref{fig:runtimes} empirically illustrates the polynomial behavior analyzed above with synthetic data. For simplicity and without loss of generality, we simulate independent white noise processes (neither serial nor cross-correlation) with two fixed lengths, $T = 10^4$ and $T = 10^5$, and for increasing the number of time series components $m$, ranging from $2$ to $30$, with increments of $2$. Additionally, the simulations were also conducted for $m = 2$ and $m = 6$, varying the time series length, $T$, from $10^2$ to $7.5\times10^6$, where the exponent is increased from $2$ to $6$ and the coefficient with increments of $2.5$. The simulation and mapping process of the resulting MTS onto an MHVG was executed on a "common" laptop with an AMD Ryzen processor with 8 cores and 16 GB of RAM.

\section{A Novel Set of Multivariate Time Series Features}\label{sec4}

Feature extraction has become a popular and manageable approach to analyzing high-dimensional time series data. Several sets of features for univariate time series have been proposed in the literature and have been made available in widely-used programming languages such as R and Python~\citep{Wang2006,hyndman2015large,fulcher2018feature,kang2020gratis,montero2020fforma,vanessa2022,bonifati2022time2feat}. However, this level of development and availability is not mirrored in the context of multivariate time series. This section aims to introduce a set of features for multivariate time series leveraging on mapping an MTS into an MHVG, encompassing: i) conventional topological features extended to MNets and ii) a novel feature specifically designed for MNets. 

\subsection{Topological Features Extended to MNets}

Common network topological features, such as node centrality, graph distances, clustering, and community, can be naturally extended to an MNet structure, including all the subgraphs mentioned in Section~\ref{subsec:mnet}. To illustrate, let us consider a local centrality measure like the \textit{degree} $k_i$ of a node $v_i$, representing the number of edges adjacent to $v_i$. In an MNet, we can compute the following three variants of $k_i$, for each layer $L_{\alpha}, \alpha =1, \ldots,m$. Here, we use the symbols $\prec$ ($\preceq$) to denote the inter-layer edges from a ``source" layer $L_{\alpha}$ (includes intra-layer edges of the ``source" layer), to a ``destination" layer $L_{\beta}$, with $\beta \neq \alpha$,
\begin{itemize}
    \item \textbf{intra-layer degree:} $k_i^{\alpha} = \sum_{j}{A_{ij}^{\alpha}}$
    \item \textbf{inter-layer degree:} $k_i^{\alpha \prec \beta} = \sum_{j}{A_{ij}^{\alpha,\beta}}$
    \item \textbf{all-layer degree:} $k_i^{\alpha \preceq \beta} = k_i^{\alpha} + k_i^{\alpha \prec \beta}$
\end{itemize}
Note that \textit{local} inter and all-layer topological measurements are asymmetric measures, meaning that $k_i^{\alpha \prec \beta} \neq k_i^{\beta \prec \alpha}$ and $k_i^{\alpha \preceq \beta} \neq k_i^{\beta \preceq \alpha}$, as the measure is relative to node-layer $v_i^{\alpha}$ or node-layer $v_i^{\beta}$.

Similarly, we can extend global topological features to MNets. Within the same example, consider the \textit{average degree} $\bar{k}$ of a graph, which calculates the arithmetic mean of the degree $k_i$ of all nodes in the graph, we can compute three variants of $\bar{k}$ in an MNet,
\begin{itemize}
    \item \textbf{average intra-degree:} $\bar{k}^{\alpha} = \frac{1}{|V_{\alpha}|} \sum_{i}{k_i^{\alpha}}$   
    \item \textbf{average inter-degree:} $\bar{k}^{\alpha,\beta} = \frac{1}{|V_{\alpha}|+|V_{\beta}|} \left( \sum_{i}{k_i^{\alpha \prec \beta}} + \sum_{j}{k_j^{\beta \prec \alpha}} \right)$
    \item \textbf{average all-degree:} $\bar{k}^{\alpha, \beta}_{all} = \frac{1}{|V_{\alpha}|+|V_{\beta}|} \left( \sum_{i}{k_i^{\alpha \preceq \beta}} + \sum_{j}{k_j^{\beta \preceq \alpha}} \right)$
\end{itemize}
Note that these features involve all elements (nodes and edges) of the corresponding (sub)graphs and are therefore symmetric features.

\newpage
In general, any common local topological feature $F_i$ can be easily extended to \textit{intra-layer features}, $F_i^{\alpha}$, by computing them over individual layers, to \textit{inter-layer features}, $F_i^{\alpha \prec \beta}$, by computing over inter-layer edges, and to \textit{all-layer features}, $F_i^{\alpha \preceq \beta}$, which compute over both intra-layer and inter-layer edges of a given node $v_i$. And, a global topological feature $\boldsymbol{F}$ can be computed in the subgraphs of the MNet: intra ($\boldsymbol{F}^{\alpha}$), inter ($\boldsymbol{F}^{\alpha, \beta}$), and all-layer graphs ($\boldsymbol{F}^{\alpha, \beta}_{all}$).

Motivated by the definition of MNet features introduced above and by the set of features proposed in~\cite{vanessa2022} for UTS analysis, namely features based on the concepts of node centrality, graph distances, clustering, and community, we propose the following topological features for each layer (or for each pair of layers) of an MNet, in particularly of an MHVGs: 
\begin{itemize}
    \item \textbf{Average degree:} The average intra-degree $\bar{k}^{\alpha}$, average inter-degree $\bar{k}^{\alpha,\beta}$ and average all-degree $\bar{k}^{\alpha, \beta}_{all}$, as formulated above.
    \item \textbf{Average path length:} Geodesic distances $d_{i, j}, i \neq j$ between node $v_i$ and $v_j$ corresponding to the length of the shortest paths between them, where the path length is the number of edges in the path. The \textit{average (intra-/inter-/all-)path length} ($\bar{d}^{\alpha}, \bar{d}^{\alpha,\beta}$ and $\bar{d}^{\alpha,\beta}_{all}$) is the arithmetic mean of the shortest paths among all pairs of nodes in (intra, inter, and all-layer) graph.
    \item \textbf{Number of communities:} The \textit{number of (intra-/inter-/all-)communities}, ($S^{\alpha}, S^{\alpha,\beta}$ and $S^{\alpha, \beta}_{all}$), is the number of groups/communities of nodes that are densely connected on the corresponding subgraph. These communities are found by performing random walks on the subgraph (intra, inter, and all-layer graph) so that short random walks tend to stay in the same community until the modularity value (defined below) cannot be increased anymore.
    \item \textbf{Modularity:} \textit{(Intra-/Inter-/All-)modularity}, ($Q^{\alpha}, Q^{\alpha,\beta}$ and $Q^{\alpha, \beta}_{all}$), measures how well a specific division of (intra-/inter-/all-)graph is into communities.
    \item \textbf{Degree distribution:} A fundamental property of a graph is the \textit{degree distribution} $P(k)$. This topological feature measures the fraction of nodes in a graph with degree $k$. Three extended variants of degree distributions are also proposed, namely, $P(k^{\alpha}), P(k^{\alpha \prec \beta})$ and $P(k^{\alpha \preceq \beta})$, in layer $L_{\alpha}, \alpha = 1, \ldots, m$, which is associated with its intra-layer degree, inter-layer degree and all-layer degree, respectively. 
    \item \textbf{Jensen–Shannon divergence:} To measure the similarity between pairs of layers in an MNet, we use the \textit{Jensen–Shannon divergence} ($JSD$) to quantify the distance between two degree distributions. $JSD^{\alpha, \beta}_{intra}, JSD_{inter}^{\alpha, \beta}$ and $JSD_{all}^{\alpha, \beta}$ measure, respectively, the $JSD$ between \textit{intra-layer}, \textit{inter-layer} and \textit{all-layer} degree distributions of layers $L_{\alpha}$ and $L_{\beta}$. For example, $JSD^{\alpha, \beta}_{intra}$ is defined as follows:
    \begin{equation}
        JSD(P(k^{\alpha})||P(k^{\beta})) = \frac{1}{2} KLD(P(k^{\alpha}) || Q(k)) + \frac{1}{2} KLD(P(k^{\beta}) || Q(k)) \nonumber
    \end{equation}
    where $Q(k) = \frac{1}{2} (P(k^{\alpha}) + P(k^{\beta}))$ and $KLD$ is the \textit{Kullback–Leibler divergence}~\footnote{Note that $JSD$ is a symmetrical version of the asymmetrical measure $KLD$.}:
    \begin{equation}
        KLD(P(k^{\alpha}) || Q(k)) = \sum_k{P(k^{\alpha}) \log_2 \left( \frac{P(k^{\alpha})}{Q(k)} \right)}. \nonumber
    \end{equation}
\end{itemize}

Table~\ref{tab:desc_features} provides additional formulation details.

\begin{table}[hbt!]
    \centering
    \footnotesize
    \caption{Summary formulation of topological features of multilayer networks.} \label{tab:desc_features}
    
    \begin{tabular}{|l|l|l|}
    \hline
        \multicolumn{1}{|c|}{\normalsize{\textbf{Feature}}}    &  \multicolumn{1}{|c|}{\normalsize{\textbf{Formulation\footnotemark[1]}}}    &   \multicolumn{1}{|c|}{\normalsize{\textbf{Note}}} \\
    \hline\hline \rule{0mm}{5ex}
    
        \multirow{3}{*}{\textbf{Average Degree}} 
            &  \(\displaystyle \bar{k}^{\alpha} = \frac{1}{N_{\alpha}} \sum_{i}{k_i^{\alpha}} \) &  \\[3ex]
            &  \(\displaystyle \bar{k}^{\alpha,\beta} = \frac{1}{N_{\alpha,\beta}} \left( \sum_{i}{k_i^{\alpha \prec \beta}} + \sum_{j}{k_j^{\beta \prec \alpha}} \right) \) &  \\[3ex]
            & \(\displaystyle \bar{k}^{\alpha,\beta}_{all} = \frac{1}{N_{\alpha,\beta}} \left( \sum_{i}{k_i^{\alpha \preceq \beta}} + \sum_{j}{k_j^{\beta \preceq \alpha}} \right) \)  &  \\[3ex]
        \hline\rule{0mm}{4ex}

        \multirow{3}{*}{\textbf{\begin{tabular}{@{}l@{}}\textbf{Degree} \\ \textbf{Distribution}\end{tabular}}}
            & \(\displaystyle P(k^{\alpha}) = \frac{n_{k^{\alpha}}}{N_{\alpha}} \)  &
            \multirow{3}{*}{\begin{tabular}{@{}l@{}}$n$: number of nodes $v_i^{\alpha}$ \\ with the corresponding \\ degree $k$ \end{tabular}} \\[3ex]
            &  \(\displaystyle P(k^{\alpha \prec \beta}) = \frac{n_{k^{\alpha \prec \beta}}}{N_{\alpha}} \)  &   \\[3ex]
            & \(\displaystyle P(k^{\alpha \preceq \beta}) = \frac{n_{k^{\alpha \preceq \beta}}}{N_{\alpha}} \)  & 
            \\[3ex]
        \hline \rule{0mm}{4ex}
        
        \multirow{3}{*}{\textbf{\begin{tabular}{@{}l@{}}\textbf{Average Path} \\ \textbf{Length}\end{tabular}}}
            & \(\displaystyle \bar{d}^{\alpha} = \frac{1}{N_{\alpha}(N_{\alpha} - 1)} \sum_{i \neq j}{d_{i,j}^{\alpha}} \)  &  \multirow{3}{*}{\begin{tabular}{@{}l@{}}$d_{i,j}$: length of the shortest \\ paths between $v_i$ and $v_j$ in \\ the corresponding subgraph \end{tabular}} \\[3ex]
            & \(\displaystyle \bar{d}^{\alpha,\beta} = \frac{1}{N_{\alpha,\beta}(N_{\alpha,\beta} - 1)} \sum_{i \neq j}{ d_{i, j}^{\alpha,\beta} } \)  &   \\[3ex]
            & \(\displaystyle \bar{d}^{\alpha,\beta}_{all} = \frac{1}{N_{\alpha,\beta}(N_{\alpha,\beta} - 1)} \sum_{i \neq j}{ d_{i,j,all}^{\alpha,\beta} } \)  & \\[3ex]
        \hline \rule{0mm}{3ex}
        
        \multirow{3}{*}{\textbf{\begin{tabular}{@{}l@{}}\textbf{Number of} \\ \textbf{Communities}\end{tabular}}}
            & \(\displaystyle S^{\alpha} = |\mathcal{C}^{\alpha}| \)  & \multirow{3}{*}{\begin{tabular}{@{}l@{}}$\mathcal{C}$: set of communities in \\ corresponding subgraph \\ \end{tabular}}  \\[2ex]
            & \(\displaystyle S^{\alpha, \beta} = |\mathcal{C}^{\alpha,\beta}| \)  &    \\[2ex]
            & \(\displaystyle S^{\alpha, \beta}_{all} = |\mathcal{C}^{\alpha,\beta}_{all}| \)  &   \\[2ex]
        \hline\rule{0mm}{4ex} 
        
        \multirow{3}{*}{\textbf{Modularity}}     
            & \(\displaystyle Q^{\alpha} = \frac{1}{2|E^{\alpha}|} \sum_{i,j}{\left[ B_{i,j} - \frac{k_i k_j}{2|E^{\alpha}|} \right]} \delta \left( c_i, c_j \right) \)  &  \(\displaystyle \boldsymbol{B} = \boldsymbol{A}^{\alpha} \)  \\[3ex]
            & \(\displaystyle Q^{\alpha, \beta} = \frac{1}{2|E^{\alpha,\beta}|} \sum_{i,j}{\left[ B_{i,j} - \frac{k_i k_j}{2|E^{\alpha,\beta}|} \right]} \delta \left( c_i, c_j \right) \)  &  $\boldsymbol{B} = \left[ \begin{smallmatrix} \boldsymbol{0} & \boldsymbol{A}^{\alpha,\beta} \\ \boldsymbol{A}^{\beta,\alpha} & \boldsymbol{0}  \end{smallmatrix} \right]$ \\[3ex]
            & \(\displaystyle Q^{\alpha, \beta}_{all} = \frac{1}{2|E^{\alpha,\beta}_{all}|} \sum_{i,j}{\left[ B_{i,j} - \frac{k_i k_j}{2|E^{
            \alpha,\beta}_{all}|} \right]} \delta \left( c_i, c_j \right) \)  &  $\boldsymbol{B} = \left[ \begin{smallmatrix} \boldsymbol{A}^{\alpha} & \boldsymbol{A}^{\alpha\beta} \\ \boldsymbol{A}^{\beta\alpha} & \boldsymbol{A}^{\beta}  \end{smallmatrix} \right]$ \\[3ex]
        \hline\rule{0mm}{4ex}
        
        \multirow{3}{*}{\textbf{\begin{tabular}{@{}l@{}}\textbf{Jensen–Shannon} \\ \textbf{Divergence}\end{tabular}}}
            & \(\displaystyle JSD_{intra}^{\alpha, \beta} = JSD(P(k^{\alpha})||P(k^{\beta})) \)  &  \\[2ex]
            & \(\displaystyle JSD_{inter}^{\alpha, \beta} = JSD(P(k^{\alpha \prec \beta})||P(k^{\beta \prec \alpha})) \)  &  \\[2ex]
            & \(\displaystyle JSD_{all}^{\alpha, \beta} =  JSD(P(k^{\alpha \preceq \beta})||P(k^{\beta \preceq \alpha})) \)  &  \\[2ex]
        \hline\rule{0mm}{4ex}
        
        \textbf{\begin{tabular}{@{}l@{}}\textbf{Average Ratio} \\ \textbf{Degree}\end{tabular}} & \(\displaystyle \bar{r}^{\alpha \preceq \beta} = \frac{1}{|N_{\alpha}|} \sum_{i}{r_i^{\alpha \preceq \beta}} \)&   \\[3ex]
        
    \hline 
    \end{tabular}
    \footnotetext[1]{Remember that $V^{\alpha}$ is the node set in layer $L_{\alpha}$, we define $N_{\alpha} = |V^{\alpha}|$ and $N_{\alpha,\beta} = |V^{\alpha}| + |V^{\beta}|$ the number of nodes in the corresponding layer(s).}
\end{table}

\subsection{Ratio Degree: a new MNet topological feature}

The \textit{ratio degree} is introduced here as a new topological feature for multilayer graph analysis, aiming to capture the relationship between intra-layer and inter-layer connections. 

For any two different layers $L_{\alpha}$ and $L_{\beta}$, $\alpha,\beta = 1, \ldots, m$, with $\alpha \neq \beta$, the ratio degree of node $v_i^{\alpha}, i = 1, \ldots, T,$ from layer $L_{\alpha}$ to layer $L_{\beta}$ is defined as 
\begin{equation}
    r_i^{\alpha \preceq \beta} = \frac{k_i^{\alpha \prec \beta}}{k_i^{\alpha}}.
\end{equation}

The \textit{average ratio degree}, $\bar{r}^{\alpha \preceq \beta}$, is the arithmetic mean of the ratio degree of all nodes of layer $L^{\alpha}$. This ratio between the external and the internal connections of a given layer provides insights into the prevalence of the Cross-HV relationships between nodes across layers compared to their horizontal visibility relationships within the layer. Thus, if $\bar{r}^{\alpha \preceq \beta} > 1$, it indicates an emphasis on the Cross-HV relation. Conversely, if $\bar{r}^{\alpha \preceq \beta} < 1$, it suggests a significance on horizontal visibility relations within the layer. If $\bar{r}^{\alpha \preceq \beta}=1$, both serial and cross-horizontal visibility relationships are similar. Note that the ratio degree and the average ratio degree are asymmetric measures. Therefore, it is not necessarily true that if $\bar{r}^{\alpha \preceq \beta} = \bar{r}^{\beta  \preceq \alpha}$, then $\bar{r}^{\alpha  \preceq \beta} = \bar{r}^{\beta  \preceq \alpha}$ also.

In this work, we will use the term \textit{relational features} to refer to similarity measures like JSD and average ratio degree.

\subsection{The MHVG Features Set}\label{sec4_3}

\begin{figure}[!ht]
\centering
\includegraphics[width=1\textwidth]{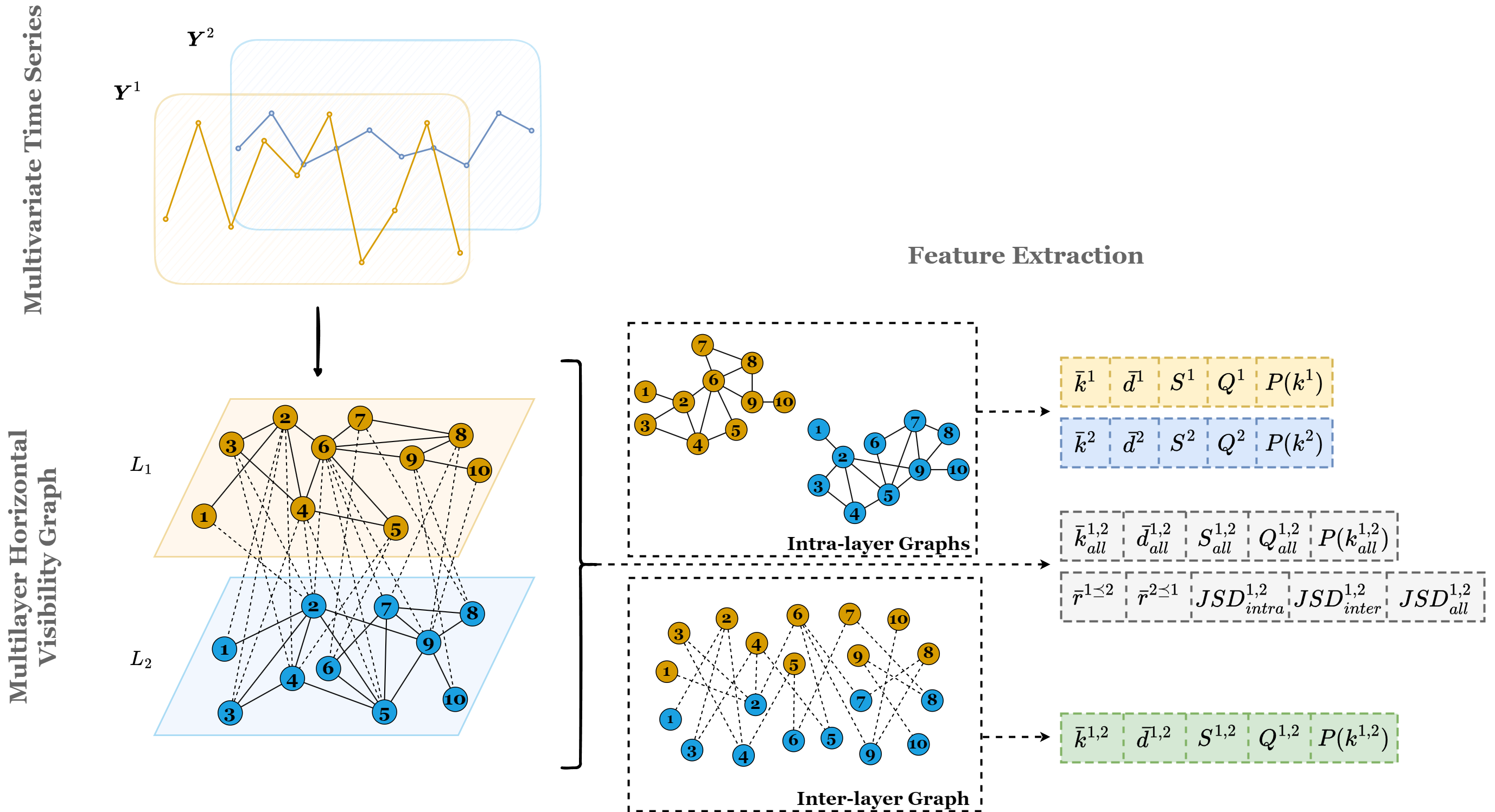}
\caption{Schematic diagram of the process for extracting the multilayer network feature set. A multivariate time series $\boldsymbol{Y}$ is mapped into a Multilayer Horizontal Visibility Graph. Global topological features ($\bar{k}, \bar{d}, S, Q$ and $P(k)$) and relational features ($\bar{r}$ and $JSD$). are computed for each of its subgraphs (intra-layer, inter-layer, and all-layer graphs).} \label{fig:process}
\end{figure}

The set of features defined above and summarized in Table~\ref{tab:desc_features} constitutes a feature set extracted from MHVG, as shown in Figure~\ref{fig:process}. This feature set is proposed for characterizing a Multivariate Time Series (MTS).

\section{Empirical evaluation of MHVG}\label{sec5}

In this section, we investigate the utility of the mapping method and the feature set introduced above for characterizing and analyzing MTS data. We also assess the applicability of the methodology and its performance in various MTS mining tasks. We begin by utilizing synthetic bivariate time series generated from bivariate time series models to control for MTS correlation properties, both serial and cross-correlation. Subsequently, we apply the methodology to real and benchmark datasets, encompassing a diverse range of characteristics and featuring data with varying dimensions, including differences in time series length and the number of time series components.

Before delving into a detailed analysis of the proposed MHVG feature set and its performance on different time series models, we provide some considerations about the methodology's implementation. Finally, we finish with the MTS mining analysis on both the synthetic and benchmark datasets to verify the applicability and advantage of introducing inter-layer edges to the horizontal visibility method.

\newpage

\subsection{Implementation Details}\label{subsec:implem}

For illustrative purposes, in Figure~\ref{fig:process}, a bivariate time series, $m=2$  was considered. However, the method is extensible to any value of $m$, as shown in the subsequent analysis of the benchmark datasets.

We follow the Algorithm~\ref{alg:mhvg} to map a multivariate time series $\boldsymbol{Y}$ into the corresponding MHVG. Specifically, we use the algorithm implemented as proposed in~\cite{Lan2015} to create the intra-layer HVGs, $\{G^{\alpha}\}_{\alpha = 1}^m$, for each time series component, $\{Y_{\alpha,t}\}, \alpha=1, \ldots, m, t = 1, \ldots, T$, and Algorithm~\ref{alg:chvg}, which follows the mapping method based on cross-horizontal visibility criteria proposed in Section~\ref{subsec:cross_hv}, to establish the inter-layer edges between the pairwise layers. We fix the subgraphs corresponding to intra-, inter-, and all-layers via the corresponding adjacency submatrices of the resulted MHVG (see Section~\ref{subsec:mnet}). Finally, we compute the corresponding topological features described in Section~\ref{sec4} using the methodologies and algorithms described below.

The \textbf{average degree ($\bar{k}$)} and \textbf{average ratio degree ($\bar{r}$)} are calculated by the arithmetic mean of the degrees $k_i$ and ratio degrees $r_i$, respectively, of all nodes $v_i$ in the respective subgraph. In this work, the \textbf{average path length ($\bar{d}$)} follows an algorithm that computes the average shortest path length between all pairs of nodes (of respective subgraphs) using a breadth-first search algorithm. To calculate the \textbf{number of communities ($S$)}, the function used makes use of the known "Louvain" algorithm that finds community structures by multi-level optimization of \textbf{modularity ($Q$)} feature (see~\citealp{blondel2008fast} for more details). The \textbf{degree distributions ($P(k)$)} and \textbf{Jensen–Shannon divergence ($JSD$)} are implemented as described in the above section. 

We used C++ and its needed set of libraries (such as \texttt{igraph} and standard libraries) to implement the data structure to store an MNet and its subgraph structures and to compute the functions to extract the topological features. For reproducibility purposes, the datasets and results are made available in \url{https://github.com/vanessa-silva/MHVG2MTS}.

\subsection{Synthetic Datasets}

We consider a set of six bivariate time series models (choosing $m=2$ for reasons of simplicity in generation and explanation for subsequent analysis), denoted as Data Generating Processes (DGPs), summarized in Table~\ref{tab:mts_setting}. These MTS models exhibit particular characteristics in terms of serial and cross-correlation (see Section~\ref{subsec:mts}), including: \textit{White Noise} (WN) processes simulating noise effects, with one process devoid of any correlation and the other featuring strong contemporaneous correlation; \textit{Vector Autoregression} (VAR) processes simulate smooth linear data, presenting both serial and cross-correlation; \textit{Vector Generalized Autoregressive Conditional Heteroskedasticity} (VGARCH) processes simulate nonlinear data with persistent periods of high or low volatility. The parameters of each DGP are chosen to ensure that the data exhibits a range of serial and cross-correlation properties, as described in Table~\ref{tab:mts_setting}. A detailed description of the DGP, their properties, and computational details are presented in  Appendix~\ref{app:mts_models}.

For each DGP in Table~\ref{tab:mts_setting}, we generated 100 instances of length $T = 10000$. As illustrated in Figure~\ref{fig:process}, we map each bivariate time series into an MHVG, highlight the intra-, inter-, and all-layer graphs, and extract the corresponding topological features. 

\begin{table}[!ht]
\centering
\caption[Summary of Data Generating Processes (DGP).]{Each synthetic dataset is generated following bivariate time series models with specified parameters and main characteristics. See Appendix~\ref{app:mts_models} for more details.}
\label{tab:mts_setting}
\begin{tabular}{lllc}
\hline
\textbf{DGP} & \textbf{Parameters} & \textbf{Characteristics} & \textbf{Notation} \\
\hline
\multirow{2}{*}{\begin{tabular}{@{}l@{}}Independent \\ White Noise\end{tabular}} & $\displaystyle \boldsymbol{\epsilon}_t \sim N(0,1)$ & \multirow{2}{*}{\begin{tabular}{@{}l@{}}Noise effect \\ No correlation\end{tabular}}  &  \texttt{iBWN} \\
 & &  & \\[1ex]
 \multirow{2}{*}{\begin{tabular}{@{}l@{}}Correlated \\ White Noise\end{tabular}} & $\displaystyle \bigl[ \begin{smallmatrix} \epsilon_{1,t} \\ \epsilon_{2,t} \end{smallmatrix} \bigr] \sim N\left(0, \bigl[ \begin{smallmatrix} 1.00 & 0.86\\ 0.86 & 1.50 \end{smallmatrix} \bigr] \right)$ &  \multirow{3}{*}{\begin{tabular}{@{}l@{}}Noise effect \\ No serial correlation \\ Cross-correlation \end{tabular}}  &  \texttt{cBWN} \\
 & &  & \\
& &  & \\[1ex]
\rule{0pt}{12pt}Weak VAR$(1)$ & $\displaystyle \boldsymbol{\varphi} = \bigl[ \begin{smallmatrix} 2.50 \\ 0.50 \end{smallmatrix} \bigr], \boldsymbol{\phi} = \bigl[ \begin{smallmatrix} 0.20 & 0.10 \\ 0.02 & 0.10 \end{smallmatrix} \bigr]$ & \multirow{2}{*}{\begin{tabular}{@{}l@{}}Weak correlation \\ (serial and cross)\end{tabular}}  &  \texttt{wVAR} \\[1ex]
& $\displaystyle \boldsymbol{\epsilon}_t \sim \bigl[ \begin{smallmatrix} 1.00 & 0.10 \\ 0.10 & 1.50 \end{smallmatrix} \bigr]$ &  & \\[1ex]
\rule{0pt}{12pt}Strong VAR$(1)$ & $\displaystyle \boldsymbol{\varphi} = \bigl[ \begin{smallmatrix} 0 \\ 0 \end{smallmatrix} \bigr], \boldsymbol{\phi} = \bigl[ \begin{smallmatrix} 0.70 & 0.02 \\ 0.30 & 0.80 \end{smallmatrix} \bigr]$ & \multirow{3}{*}{\begin{tabular}{@{}l@{}}Strong correlation \\ (serial and cross, lagged \\ and contemporaneous)\end{tabular}}  &  \texttt{sVAR} \\[1ex]
 & $\displaystyle \boldsymbol{\epsilon}_t \sim \bigl[ \begin{smallmatrix} 1.00 & 0.86 \\ 0.86 & 1.50 \end{smallmatrix} \bigr]$ &  & \\
  & &  \\[1ex]
\multirow{2}{*}{\begin{tabular}{@{}l@{}}Weak \\ VGARCH$(1,1)$\end{tabular}} & $\displaystyle\boldsymbol{\omega} = \bigl[ \begin{smallmatrix} 0.05 \\ 0.02 \end{smallmatrix} \bigr]$, $\boldsymbol{\alpha} = \bigl[ \begin{smallmatrix} 0.10 & 0.00 \\ 0.00 & 0.05 \end{smallmatrix} \bigr] $ & \multirow{2}{*}{\begin{tabular}{@{}l@{}}No serial correlation \\ Weak cross-correlation\end{tabular}} &  \texttt{wGARCH} \\[1ex]
& $\displaystyle \boldsymbol{\beta} = \bigl[ \begin{smallmatrix} 0.85 & 0.00 \\ 0.00 & 0.88 \end{smallmatrix} \bigr], \boldsymbol{\epsilon}_t \sim \bigl[ \begin{smallmatrix} 1.00 & 0.10 \\ 0.10 & 1.50 \end{smallmatrix} \bigr]$ &  & \\[1ex]
\multirow{2}{*}{\begin{tabular}{@{}l@{}}Strong \\ VGARCH$(1,1)$\end{tabular}} & $\displaystyle \boldsymbol{\omega} = \bigl[ \begin{smallmatrix} 0.05 \\ 0.02 \end{smallmatrix} \bigr]$, $\boldsymbol{\alpha} = \bigl[ \begin{smallmatrix} 0.10 & 0.00 \\ 0.00 & 0.05 \end{smallmatrix} \bigr] $ & \multirow{2}{*}{\begin{tabular}{@{}l@{}}Strong contemporaneous \\ cross-correlation\end{tabular}} &  \texttt{sGARCH} \\[1ex]
 & $\displaystyle \boldsymbol{\beta} = \bigl[ \begin{smallmatrix} 0.85 & 0.00 \\ 0.00 & 0.88 \end{smallmatrix} \bigr], \boldsymbol{\epsilon}_t \sim \bigl[ \begin{smallmatrix} 1.00 & 0.86 \\ 0.86 & 1.50 \end{smallmatrix} \bigr]$ &   & \\[1ex]
\hline
\end{tabular}
\end{table}

\begin{figure}[H]
\centering
\includegraphics[width=1\textwidth]{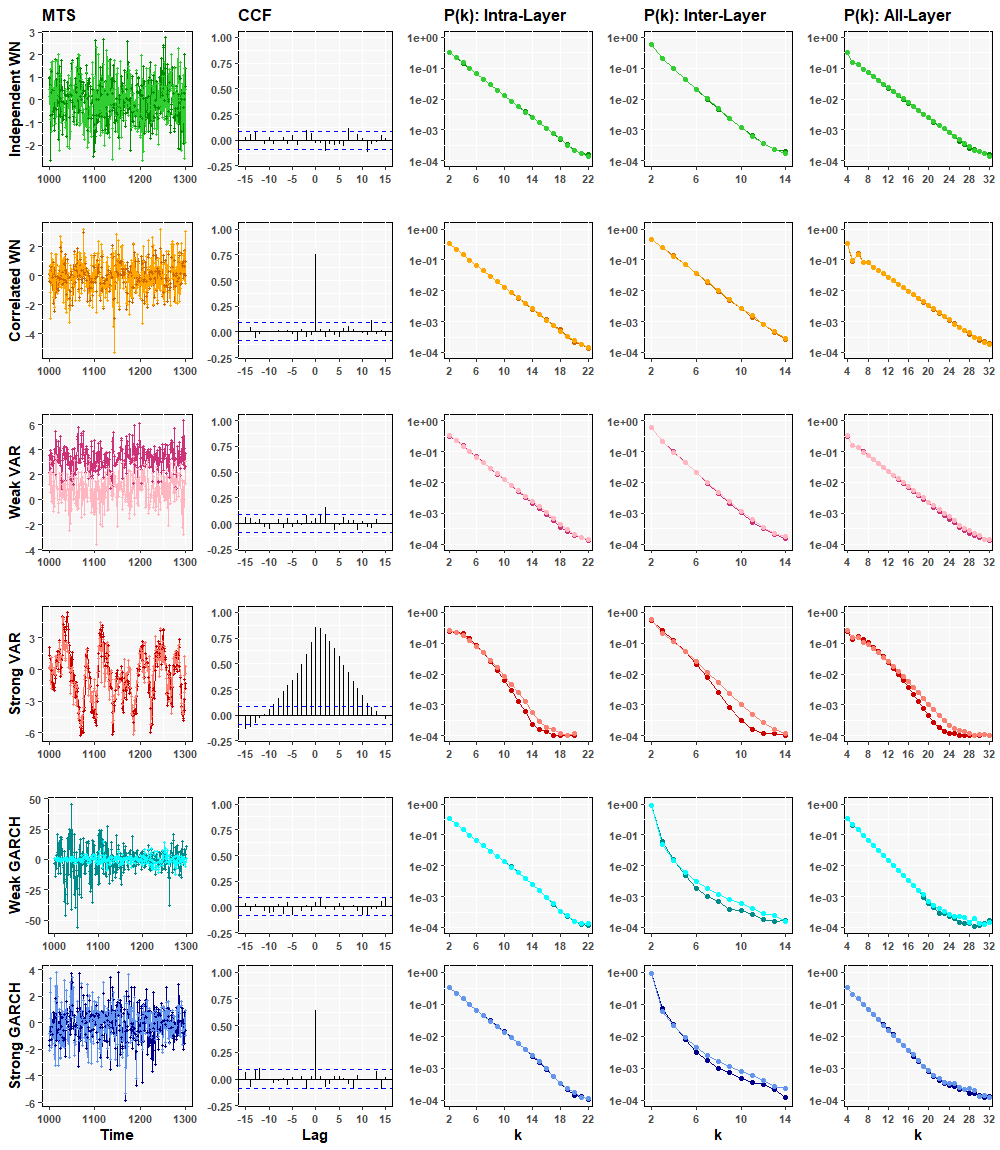}
\caption{Analysis plots of DGP. The first column shows a subset of timestamps of each DGP, the second plots the cross-correlation function between their corresponding time series components, and the last three show the degree distribution of the MHVGs. The plots corresponding to the degree distributions are on a semi-logarithmic scale. Lines of different colors refer to each DGP model ($\boldsymbol{Y}$), where the darkest colors refer to their first components ($\boldsymbol{Y}^1$) and the lighter ones to the second ($\boldsymbol{Y}^2$).} \label{fig:mts_ccf_dd}
\end{figure}

\newpage
To illustrate the procedure, we represent in Figure~\ref{fig:mts_ccf_dd} one instance with 300 observations of each DGP and the corresponding cross-correlation (CCF) plot (first two columns of the plot), the intra-, inter-, and all-layers degree distributions on a semi-logarithmic scale (last three columns of the plot). These degree distributions are computed as the arithmetic mean of the degree distributions of the 100 simulated instances.  

The plots presented in Figure~\ref{fig:mts_ccf_dd} clearly show that the degree distributions are different across the DGPs. In fact, \cite{Luque2009} has shown that the intra-layer degree distribution for white noise (uncorrelated data) follows a power law $\left(P(k) = \frac{1}{3}\left( \frac{2}{3}\right)^{k-2} \right)$ and our results indicate that strong serial correlation leads to intra-layer degree distributions that are positively skewed: as illustrated in Appendix~\ref{app:mts_models}, the \texttt{sVAR} is the only DGP that produces data with strong serial correlation. The degree distribution for the inter-layer subgraphs does not have an algebraic close form even in the simplest case of two uncorrelated white noises. However, extensive simulations indicate that it does not follow the power law $P(k) = \frac{1}{3}\left( \frac{2}{3}\right)^{k-2} $, as illustrated in the first line,  third column of Figure~\ref{fig:mts_ccf_dd}. The plots also indicate that inter-layer degree distribution depends both on the correlation between the two time series (CCF represented in the second column of the plot in Figure~\ref{fig:mts_ccf_dd}) and the serial correlation within each time series. Moreover, we note that inter-layer degree distributions for \texttt{sVAR} are positively skewed, for VGARCH models, \texttt{wGARCH} and \texttt{sGARCH}, are exponentially shaped while the remaining are approximately linear. Once again, a slower decay of the lagged correlation leads to a longer tail in the degree distribution. Also, the degree distribution curves corresponding to the VGARCH models stand out from the others, especially the inter- and all-layer degree distributions. The exponential shape of the inter-layer degree distributions is induced by the heteroscedasticity and volatility clusters in the data which limit cross-horizontal visibility to the nearest neighbors. 

\subsection{MTS Feature Set via MHVG}

The results for all the 21 features introduced in Section~\ref{sec4} and all DGPs, organized by subgraph structure, are summarized, mean (standard deviation), in Tables~\ref{tab:dgp_subgraphs} and \ref{tab:dgp_allrelat}. The values have been Min-Max normalized (across models) for comparison purposes since the range of the different features varies across the different DGPs. The cells in the tables are colored with a gradient based on the mean values: cells with a maximum value of 1 are colored red, cells with a minimum value of 0 are colored white, and the remainder with a hue of red color proportional to its value.

The results indicate that each set of features --- intra-layer (first two columns of each feature in Table~\ref{tab:dgp_subgraphs}), inter-layer (third column of each feature in Table~\ref{tab:dgp_subgraphs}), all-layer (top of Table~\ref{tab:dgp_allrelat}) and relational (bottom of Table~\ref{tab:dgp_allrelat}) --- distinguishes two groups of MTS based on properties related to correlation (serial and cross) and volatility clustering.

Focusing on the characteristic of conditional heteroscedasticity, we can observe that the proposed mapping is sensitive to this property, particularly the Cross-HVG mapping. The visibility criterion is dependent on the very high and very low values of the data over time. Therefore, the heteroscedasticity and correlation between data impose additional limits on the cross visibility between values of different variables over time, leading to greater variability in the topological features of the HVG and, especially,  Cross-HVG. This observation is evident in Table~\ref{tab:dgp_subgraphs} where the standard deviation values corresponding to the 100 samples of each DGP analyzed are higher for the VGARCH models. 

\newpage

\newpage
\begin{table}[!hb]
\small
    \caption[Topological global features of intra-layer graphs and inter-layer graph of MHVGs from DGP]{Mean values (standard deviation) of the $100$ instances of each DGP for each topological global feature from intra-layer graphs, $G^1$ and $G^2$, and inter-layer graph, $G^{1,2}$, resulting from the corresponding MHVGs.}
    \label{tab:dgp_subgraphs}
    \centering

\begin{tabular}{|>{\centering\arraybackslash}p{1.5cm}|>{\centering\arraybackslash}p{1cm}|>{\centering\arraybackslash}p{1cm}|>{\centering\arraybackslash}p{1cm}|>{\centering\arraybackslash}p{1cm}|>{\centering\arraybackslash}p{1cm}|>{\centering\arraybackslash}p{1cm}|}

\hline
\multirow{2}{*}{\textcolor{black}{\textbf{ DGP }}}
 & \multicolumn{3}{c|}{\textcolor{black}{\textbf{Average Degree}}} 
 & \multicolumn{3}{c|}{\textcolor{black}{\textbf{Average Path Length}}}  \\
 
 & \multicolumn{1}{c}{\textcolor{black}{\textbf{\footnotesize{$\bar{k}^1$}}}} 
 & \multicolumn{1}{c}{\textcolor{black}{\textbf{\footnotesize{$\bar{k}^2$}}}}
 & \multicolumn{1}{c|}{\textcolor{black}{\textbf{\footnotesize{$\bar{k}^{1,2}$}}}}
 & \multicolumn{1}{c}{\textcolor{black}{\textbf{\footnotesize{$\bar{d}^1$}}}} 
 & \multicolumn{1}{c}{\textcolor{black}{\textbf{\footnotesize{$\bar{d}^2$}}}} 
 & \multicolumn{1}{c|}{\textcolor{black}{\textbf{\footnotesize{$\bar{d}^{1,2}$}}}} \\[1ex]
\hline

\multirow{2}{*}{\texttt{iBWN}} 
 & \cellcolor{ff0000}{0.805}& \cellcolor{ff0202}{0.756} & \cellcolor{ff6262}{0.615} 
 & \cellcolor{ffffff}{0.044}& \cellcolor{ffffff}{0.049} & \cellcolor{fff9f9}{0.012} \\
 & \cellcolor{ff0000}{\tiny{(0.081)}} & \cellcolor{ff0202}{\tiny{(0.077)}} & \cellcolor{ff6262}{\tiny{(0.033)}}
 & \cellcolor{ffffff}{\tiny{(0.019)}} & \cellcolor{ffffff}{\tiny{(0.023)}} & \cellcolor{fff9f9}{\tiny{(0.024)}} \\
\hline
\multirow{2}{*}{\texttt{cBWN}} 
 & \cellcolor{ff0303}{0.802}& \cellcolor{ff0606}{0.752} & \cellcolor{ff0000}{0.940}
 & \cellcolor{fffefe}{0.045}& \cellcolor{fffefe}{0.050} & \cellcolor{ffffff}{0.007} \\
 & \cellcolor{ff0303}{\tiny{(0.083)}} & \cellcolor{ff0606}{\tiny{(0.080)}} & \cellcolor{ff0000}{\tiny{(0.078)}}
 & \cellcolor{fffefe}{\tiny{(0.022)}} & \cellcolor{fffefe}{\tiny{(0.022)}} & \cellcolor{ffffff}{\tiny{(0.015)}} \\
\hline

\multirow{2}{*}{\texttt{wVAR}} 
 & \cellcolor{ff0e0e}{0.790} & \cellcolor{ff0000}{0.759} & \cellcolor{ff6363}{0.615} 
 & \cellcolor{fff6f6}{0.058} & \cellcolor{fff9f9}{0.056} & \cellcolor{fffcfc}{0.009} \\
 & \cellcolor{ff0e0e}{\tiny{(0.079)}} &\cellcolor{ff0000}{\tiny{(0.090)}} & \cellcolor{ff6363}{\tiny{(0.033)}} 
 & \cellcolor{fff6f6}{\tiny{(0.022)}} & \cellcolor{fff9f9}{\tiny{(0.025)}} & \cellcolor{fffcfc}{\tiny{(0.016)}} \\
\hline
\multirow{2}{*}{\texttt{sVAR}} 
 & \cellcolor{ffeaea}{0.561} & \cellcolor{ff9e9e}{0.601} & \cellcolor{ff4e4e}{0.683} 
 & \cellcolor{ff0000}{0.449} & \cellcolor{ff2b2b}{0.328} & \cellcolor{fffbfb}{0.011}  \\
 & \cellcolor{ffeaea}{\tiny{(0.121)}} & \cellcolor{ff9e9e}{\tiny{(0.108)}} & \cellcolor{ff4e4e}{\tiny{(0.092)}} 
 & \cellcolor{ff0000}{\tiny{(0.050)}} & \cellcolor{ff2b2b}{\tiny{(0.039)}} & \cellcolor{fffbfb}{\tiny{(0.025)}} \\
\hline

\multirow{2}{*}{\texttt{wGARCH}} 
 & \cellcolor{ffffff}{0.540} & \cellcolor{ffcece}{0.554} & \cellcolor{ffffff}{0.102} 
 & \cellcolor{ff2a2a}{0.380} & \cellcolor{ff2d2d}{0.325} & \cellcolor{ff0000}{0.252} \\
 & \cellcolor{ffffff}{\tiny{(0.159)}} & \cellcolor{ffcece}{\tiny{(0.140)}} & \cellcolor{ffffff}{\tiny{(0.135)}} 
 & \cellcolor{ff2a2a}{\tiny{(0.136)}} & \cellcolor{ff2d2d}{\tiny{(0.097)}} & \cellcolor{ff0000}{\tiny{(0.188)}} \\
\hline
\multirow{2}{*}{\texttt{sGARCH}} 
 & \cellcolor{fffcfc}{0.542} & \cellcolor{ffffff}{0.505} & \cellcolor{fff1f1}{0.146} 
 & \cellcolor{ff2525}{0.390} & \cellcolor{ff0000}{0.385} & \cellcolor{ff1414}{0.232} \\
 & \cellcolor{fffcfc}{\tiny{(0.184)}} & \cellcolor{ffffff}{\tiny{(0.186)}} & \cellcolor{fff1f1}{\tiny{(0.174)}} 
 & \cellcolor{ff2525}{\tiny{(0.138)}} & \cellcolor{ff0000}{\tiny{(0.149)}} & \cellcolor{ff1414}{\tiny{(0.217)}} \\

\hline

\end{tabular}

\medskip

\begin{tabular}{|>{\centering\arraybackslash}p{1.5cm}|>{\centering\arraybackslash}p{1cm}|>{\centering\arraybackslash}p{1cm}|>{\centering\arraybackslash}p{1cm}|>{\centering\arraybackslash}p{1cm}|>{\centering\arraybackslash}p{1cm}|>{\centering\arraybackslash}p{1cm}|}

\hline  
\multirow{2}{*}{\textcolor{black}{\textbf{ DGP }}}
 & \multicolumn{3}{c|}{\textcolor{black}{\textbf{Num. of Communities}}} 
 & \multicolumn{3}{c|}{\textcolor{black}{\textbf{Modularity}}} \\
 
 & \multicolumn{1}{c}{\textcolor{black}{\textbf{\footnotesize{$S^1$}}}}
 & \multicolumn{1}{c}{\textcolor{black}{\textbf{\footnotesize{$S^2$}}}}
 & \multicolumn{1}{c|}{\textcolor{black}{\textbf{\footnotesize{$S^{1,2}$}}}} 
 & \multicolumn{1}{c}{\textcolor{black}{\textbf{\footnotesize{$Q^1$}}}}
 & \multicolumn{1}{c}{\textcolor{black}{\textbf{\footnotesize{$Q^2$}}}}  
 & \multicolumn{1}{c|}{\textcolor{black}{\textbf{\footnotesize{$Q^{1,2}$}}}} \\[1ex]
\hline

\multirow{2}{*}{\texttt{iBWN}} 
 & \cellcolor{fff0f0}{0.265} & \cellcolor{ffeded}{0.312} & \cellcolor{ffdbdb}{0.206} 
 & \cellcolor{fffefe}{0.150}& \cellcolor{fffafa}{0.207} & \cellcolor{ffb7b7}{0.319} \\
 & \cellcolor{fff0f0}{\tiny{(0.083)}} & \cellcolor{ffeded}{\tiny{(0.083)}} & \cellcolor{ffdbdb}{\tiny{(0.064)}} 
 & \cellcolor{fffefe}{\tiny{(0.055)}}& \cellcolor{fffafa}{\tiny{(0.056)}} & \cellcolor{ffb7b7}{\tiny{(0.093)}} \\
\hline
\multirow{2}{*}{\texttt{cBWN}} 
 & \cellcolor{fff3f3}{0.260}& \cellcolor{ffeeee}{0.310}& \cellcolor{ffffff}{0.126} 
 & \cellcolor{ffffff}{0.150}& \cellcolor{ffffff}{0.196} & \cellcolor{ffffff}{0.181} \\
 & \cellcolor{fff3f3}{\tiny{(0.084)}} & \cellcolor{ffeeee}{\tiny{(0.090)}}& \cellcolor{ffffff}{\tiny{(0.080)}} 
 & \cellcolor{ffffff}{\tiny{(0.051)}}& \cellcolor{ffffff}{\tiny{(0.062)}} & \cellcolor{ffffff}{\tiny{(0.141)}} \\
\hline

\multirow{2}{*}{\texttt{wVAR}} 
 & \cellcolor{ffcdcd}{0.342} & \cellcolor{ffdcdc}{0.338} & \cellcolor{ffd8d8}{0.211} 
 & \cellcolor{ffd0d0}{0.287} & \cellcolor{ffdfdf}{0.277} & \cellcolor{ffbcbc}{0.308} \\
 & \cellcolor{ffcdcd}{\tiny{(0.093)}} & \cellcolor{ffdcdc}{\tiny{(0.100)}} & \cellcolor{ffd8d8}{\tiny{(0.062)}} 
 & \cellcolor{ffd0d0}{\tiny{(0.058)}} & \cellcolor{ffdfdf}{\tiny{(0.062)}} & \cellcolor{ffbcbc}{\tiny{(0.099)}} \\
\hline
\multirow{2}{*}{\texttt{sVAR}} 
 & \cellcolor{ff0000}{0.791} & \cellcolor{ff0000}{0.700} & \cellcolor{ffe0e0}{0.195} 
 & \cellcolor{ff0000}{0.893}  & \cellcolor{ff0000}{0.857} & \cellcolor{ffd4d4}{0.263} \\
 & \cellcolor{ff0000}{\tiny{(0.104)}} & \cellcolor{ff0000}{\tiny{(0.121)}} & \cellcolor{ffe0e0}{\tiny{(0.069)}} 
 & \cellcolor{ff0000}{\tiny{(0.046)}}  & \cellcolor{ff0000}{\tiny{(0.064)}} & \cellcolor{ffd4d4}{\tiny{(0.103)}} \\
\hline

\multirow{2}{*}{\texttt{wGARCH}} 
 & \cellcolor{fffcfc}{0.239} & \cellcolor{ffffff}{0.282} & \cellcolor{ff0000}{0.696} 
 & \cellcolor{fff2f2}{0.185}  & \cellcolor{fffafa}{0.207} & \cellcolor{ff0000}{0.669} \\
 & \cellcolor{fffcfc}{\tiny{(0.104)}} & \cellcolor{ffffff}{\tiny{(0.083)}} & \cellcolor{ff0000}{\tiny{(0.212)}} 
 & \cellcolor{fff2f2}{\tiny{(0.057)}}  & \cellcolor{fffafa}{\tiny{(0.081)}} & \cellcolor{ff0000}{\tiny{(0.210)}}\\
\hline
\multirow{2}{*}{\texttt{sGARCH}} 
 & \cellcolor{ffffff}{0.234} & \cellcolor{fff4f4}{0.300} & \cellcolor{ff0b0b}{0.670} 
 & \cellcolor{fff5f5}{0.179} & \cellcolor{fff8f8}{0.212} & \cellcolor{ff1616}{0.626} \\
 & \cellcolor{ffffff}{\tiny{(0.103)}} & \cellcolor{fff4f4}{\tiny{(0.105)}} & \cellcolor{ff0b0b}{\tiny{(0.220)}} 
 & \cellcolor{fff5f5}{\tiny{(0.065)}} & \cellcolor{fff8f8}{\tiny{(0.065)}} & \cellcolor{ff1616}{\tiny{(0.240)}} \\

\hline

\end{tabular}
\end{table}

\begin{figure}[!ht]
\centering
\includegraphics[scale = 0.43]{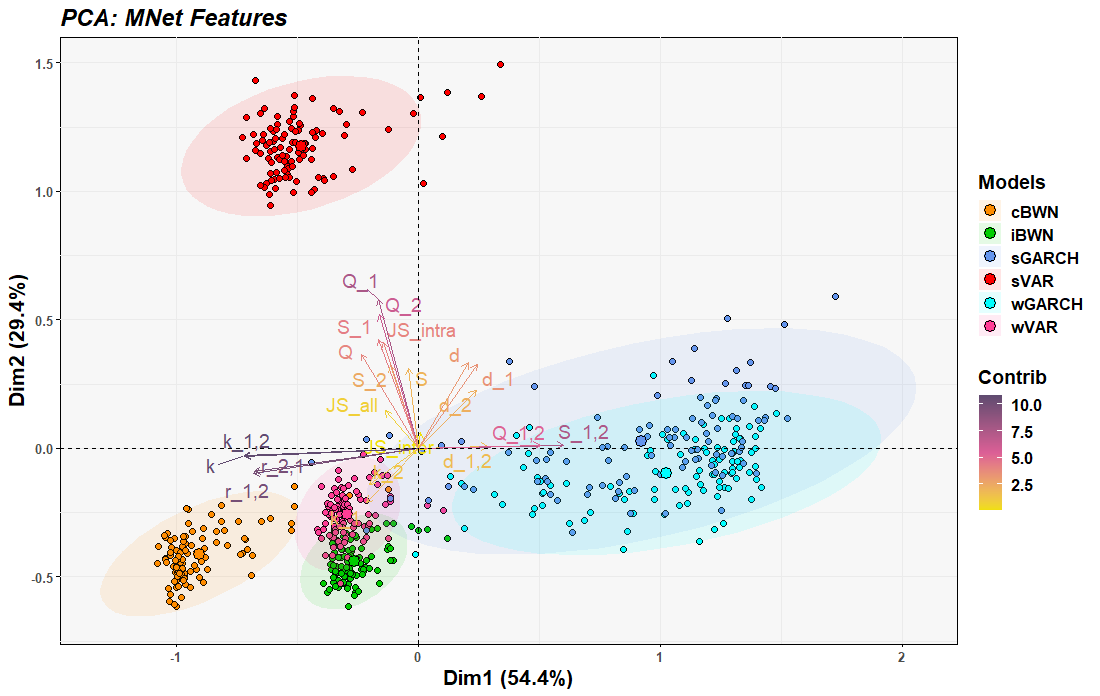}
\caption{Bi-plot of the first two principal components (PC) of principal component analysis for the Data Generating Process (DGP). Each DGP is represented by a different color and the arrows represent the contributions of the MNet features to the PCs. The larger the size, sharpness, and closeness to the red the greater the contribution of the feature. Features placed together are positively correlated and those placed on opposite quadrants are negatively correlated.} \label{fig:pcaplot}
\end{figure}

To understand which MNet topological features capture the specific properties of the MTS models, we perform PCA on the feature space. Figure~\ref{fig:pcaplot} represents a bi-plot obtained using the intra-, inter-, all-layer, and relational features, with the two principal components (PC) explaining $83.8\%$ of the variance. The bi-plots resulting from PCA in restricted feature sets are represented in  Figure~\ref{fig:pcaplot_sets} in Appendix~\ref{app:mts_models}). Overall, we can say that the average degree and average ratio degree, $\bar{k}$ and $\bar{r}$, are positively and negatively correlated, respectively, with the average path length, $\bar{d}$. The community-related features of the intra- and all-layer graphs are positively correlated but less correlated with the community-related features of the inter-layer graphs. The features that most contribute to the first two PCs are the $\bar{k}^{1,2}, S^{1,2}$ and $Q^{1,2}$ of the inter-layer graphs, the $\bar{k}^{1, 2}_{all}$ of the all-layer graphs, the $\bar{r}^{1 \preceq 2}$ and $\bar{r}^{2 \preceq 1}$ of the relational layers, and $Q^1$ and $Q^2$ of the intra-layer graphs (see Figure~\ref{fig:barplot} of Appendix~\ref{app:mts_models}). 
Figure~\ref{fig:pcaplot} clearly shows four groups of models, VGARCH, \texttt{sVAR}, \texttt{cBWN}, and a group constituted by \texttt{wVAR} and \texttt{iBWN}, identifying the topological features that characterize them.

\newpage
\begin{table}[!htb]
\small
    \caption[Topological global and relational metrics of all-layer graph of MHVGs from DGP]{Mean values (standard deviation)  for each topological global and relational features from all-layer graphs, $G^{1,2}_{all}$, resulting from MHVGs of DGP, computed over the $100$ instances of each DGP.}
    \label{tab:dgp_allrelat}
    \centering

\newpage

\begin{tabular}{|>{\centering\arraybackslash}p{1.5cm}|>{\centering\arraybackslash}p{2cm}|>{\centering\arraybackslash}p{2cm}|>{\centering\arraybackslash}p{2cm}|>{\centering\arraybackslash}p{2cm}|}

\hline  
\multirow{3}{*}{\textcolor{black}{\textbf{ DGP }}} 
& \textcolor{black}{\textbf{Average}} 
& \textcolor{black}{\textbf{Average}} 
& \textcolor{black}{\textbf{Num. of}} 
& \multirow{2}{*}{\textcolor{black}{\textbf{Modularity}}} \\
 
& \textcolor{black}{\textbf{Degree}} 
& \textcolor{black}{\textbf{Path Length}}
& \textcolor{black}{\textbf{Communities}} 
& \\

& \multicolumn{1}{c|}{\textcolor{black}{\textbf{\footnotesize{$\bar{k}^{1,2}_{all}$}}}} 
& \multicolumn{1}{c|}{\textcolor{black}{\textbf{\footnotesize{$\bar{d}^{1,2}_{all}$}}}} 
& \multicolumn{1}{c|}{\textcolor{black}{\textbf{\footnotesize{$S^{1,2}_{all}$}}}}
& \multicolumn{1}{c|}{\textcolor{black}{\textbf{\footnotesize{$Q^{1,2}_{all}$}}}} \\[1ex]
\hline

\multirow{2}{*}{\texttt{iBWN}} 
 & \cellcolor{ff6262}{0.617} & \cellcolor{ffffff}{0.042} & \cellcolor{ffffff}{0.237} & \cellcolor{ffcaca}{0.338}  \\
 & \cellcolor{ff6262}{\tiny{(0.033)}} & \cellcolor{ffffff}{\tiny{(0.018)}} & \cellcolor{ffffff}{\tiny{(0.107)}} & \cellcolor{ffcaca}{\tiny{(0.052)}}  \\
\hline
\multirow{2}{*}{\texttt{cBWN}} 
 & \cellcolor{ff0000}{0.940} & \cellcolor{fff9f9}{0.051} & \cellcolor{ffa9a9}{0.382} & \cellcolor{ff8686}{0.507}  \\
 & \cellcolor{ff0000}{\tiny{(0.078)}} & \cellcolor{fff9f9}{\tiny{(0.018)}} & \cellcolor{ffa9a9}{\tiny{(0.087)}} & \cellcolor{ff8686}{\tiny{(0.064)}}   \\
\hline

\multirow{2}{*}{\texttt{wVAR}} 
 & \cellcolor{ff6262}{0.616} & \cellcolor{fff7f7}{0.054} & \cellcolor{ffd7d7}{0.305} & \cellcolor{ffacac}0.413\\
 & \cellcolor{ff6262}{\tiny{(0.033)}} & \cellcolor{fff7f7}{\tiny{(0.022)}} & \cellcolor{ffd7d7}{\tiny{(0.107)}} & \cellcolor{ffacac}\tiny{(0.049)}  \\
\hline
\multirow{2}{*}{\texttt{sVAR}} 
 & \cellcolor{ff4e4e}{0.682} & \cellcolor{ff0000}{0.457} &  \cellcolor{ff0000}{0.457} & \cellcolor{ff0000}0.842 \\
 & \cellcolor{ff4e4e}{\tiny{(0.092)}} & \cellcolor{ff0000}{\tiny{(0.04955)}} & \cellcolor{ff0000}{\tiny{(0.127)}} & \cellcolor{ff0000}\tiny{(0.05673)} \\
\hline

\multirow{2}{*}{\texttt{wGARCH}} 
 & \cellcolor{ffffff}{0.104} & \cellcolor{ff6262}{0.297} & \cellcolor{ffd3d3}{0.310} & \cellcolor{ffffff}0.206 \\
 & \cellcolor{ffffff}{\tiny{(0.134)}} & \cellcolor{ff6262}{\tiny{(0.065)}} & \cellcolor{ffd3d3}{\tiny{(0.108)}} & \cellcolor{ffffff}\tiny{(0.076)}  \\
\hline
\multirow{2}{*}{\texttt{sGARCH}} 
 & \cellcolor{fff1f1}{0.147} & \cellcolor{ff2a2a}{0.388} & \cellcolor{ff8c8c}{0.431} & \cellcolor{ffb3b3}0.3954 \\
 & \cellcolor{fff1f1}{\tiny{(0.173)}} & \cellcolor{ff2a2a}{\tiny{(0.120)}} & \cellcolor{ff8c8c}{\tiny{(0.118)}} & \cellcolor{ffb3b3}\tiny{(0.088)} \\

\hline

\end{tabular}

\medskip

\begin{tabular}{|>{\centering\arraybackslash}p{1.5cm}|>{\centering\arraybackslash}p{1.5cm}|>{\centering\arraybackslash}p{1.5cm}|>{\centering\arraybackslash}p{1.5cm}|>{\centering\arraybackslash}p{1.55cm}|>{\centering\arraybackslash}p{1.55cm}|}

\hline  
\multirow{3}{*}{\textcolor{black}{\textbf{ DGP }}} 
& \multicolumn{2}{c|}{\textcolor{black}{\textbf{Average Ratio Degree}}}  
& \multicolumn{3}{c|}{\textcolor{black}{\textbf{Jensen–Shannon Divergence}}} \\

& \multicolumn{1}{c}{\textcolor{black}{\textbf{\footnotesize{$\bar{r}^{1 \preceq 2}$}}}} 
& \multicolumn{1}{c|}{\textcolor{black}{\textbf{\footnotesize{$\bar{r}^{2 \preceq 1}$}}}}
& \multicolumn{1}{c}{\textcolor{black}{\textbf{\footnotesize{$JSD^{\alpha,\beta}_{intra}$}}}} 
& \multicolumn{1}{c}{\textcolor{black}{\textbf{\footnotesize{$JSD^{\alpha,\beta}_{inter}$}}}} 
& \multicolumn{1}{c|}{\textcolor{black}{\textbf{\footnotesize{$JSD^{\alpha,\beta}_{all}$}}}} \\[1ex]
\hline

\multirow{2}{*}{\texttt{iBWN}} 
 & \cellcolor{ff6c6c}{0.586} & \cellcolor{ff6a6a}{0.582} & \cellcolor{fff5f5}{0.170} & \cellcolor{fffefe}{0.034} & \cellcolor{ffefef}0.090 \\
 & \cellcolor{ff6c6c}{\tiny{(0.028)}} & \cellcolor{ff6a6a}{\tiny{(0.037)}} & \cellcolor{fff5f5}{\tiny{(0.071)}} & \cellcolor{fffefe}{\tiny{(0.0355)}} & \cellcolor{ffefef}{\tiny{(0.047)}} \\
\hline
\multirow{2}{*}{\texttt{cBWN}} 
 & \cellcolor{ff0000}{0.942} & \cellcolor{ff0000}{0.931} & \cellcolor{fff7f7}{0.166} & \cellcolor{ffaeae}{0.061} & \cellcolor{ff4949}0.244 \\
 & \cellcolor{ff0000}{\tiny{(0.071)}} & \cellcolor{ff0000}{\tiny{(0.084)}} & \cellcolor{fff7f7}{\tiny{(0.079)}} & \cellcolor{ffaeae}{\tiny{(0.076)}} & \cellcolor{ff4949}{\tiny{(0.236)}} \\
\hline

\multirow{2}{*}{\texttt{wVAR}} 
 & \cellcolor{ff7272}{0.565} & \cellcolor{ff6e6e}{0.571} & \cellcolor{ffe3e3}{0.207} & \cellcolor{ffffff}{0.034} & \cellcolor{ffe8e8}{0.096} \\
 & \cellcolor{ff7272}{\tiny{(0.032)}} & \cellcolor{ff6e6e}{\tiny{(0.034)}} & \cellcolor{ffe3e3}{\tiny{(0.074)}} & \cellcolor{ffffff}{\tiny{(0.041)}} & \cellcolor{ffe8e8}{\tiny{(0.051)}} \\
\hline
\multirow{2}{*}{\texttt{sVAR}} 
 & \cellcolor{ff7070}{0.574} & \cellcolor{ff6b6b}{0.579} & \cellcolor{ff0000}{0.682} & \cellcolor{ff0000}{0.120} & \cellcolor{ff0000}{0.314} \\
 & \cellcolor{ff7070}{\tiny{(0.106)}} & \cellcolor{ff6b6b}{\tiny{(0.091)}} & \cellcolor{ff0000}{\tiny{(0.128)}} & \cellcolor{ff0000}{\tiny{(0.148)}} & \cellcolor{ff0000}{\tiny{(0.225)}} \\
\hline

\multirow{2}{*}{\texttt{wGARCH}} 
 & \cellcolor{ffffff}{0.103} & \cellcolor{ffffff}{0.097} & \cellcolor{fffdfd}{0.154} & \cellcolor{ff6767}{0.085} & \cellcolor{ffffff}{0.075} \\
 & \cellcolor{ffffff}{\tiny{(0.136)}} & \cellcolor{ffffff}{\tiny{(0.127)}} & \cellcolor{fffdfd}{\tiny{(0.065)}} & \cellcolor{ff6767}{\tiny{(0.110)}} & \cellcolor{ffffff}{\tiny{(0.065)}} \\
\hline
\multirow{2}{*}{\texttt{sGARCH}} 
 & \cellcolor{fff1f1}{0.149} & \cellcolor{fff0f0}{0.145} & \cellcolor{ffffff}{0.149} & \cellcolor{ff1212}{0.114} & \cellcolor{ffbcbc}{0.137} \\
 & \cellcolor{fff1f1}{\tiny{(0.179)}} & \cellcolor{fff0f0}{\tiny{(0.170)}} & \cellcolor{ffffff}{\tiny{(0.077)}} & \cellcolor{ff1212}{\tiny{(0.161)}} & \cellcolor{ffbcbc}{\tiny{(0.172)}} \\
 \hline

\end{tabular}
\end{table}

The strong ACF and CCF of the \texttt{sVAR} are represented by high values for the number of communities and modularity in its intra- and all-layer graphs. Inter-layer graphs present higher values of community-related features for VGARCH models. The average path length represents the VGARCH models, in particular, the average path length of the all-layer graphs tries to distinguish both \texttt{wGARCH} and \texttt{sGARCH}. The strong contemporaneous CCF of \texttt{cBWN} is represented by high values of average ratio degree features, such as the average degree values of its inter and all-layer graphs. \texttt{iBWN} and \texttt{wVAR} are represented by high values of intra-layer average degree.

The above results indicate that the topological features extracted from MHVG are adequate as a set of MTS features. To further explore this idea, in the next section, we proceed with multivariate time series mining tasks of the DGPs and the benchmark datasets via MNet topological features.

\newpage

\subsection{Mining Time Series with MNet Features} \label{subsec:clust}

To conclude our analysis and demonstrate the applicability and practicality of the proposed method, this section presents two direct applications in one of the most common time series mining tasks. We begin with a multivariate time series clustering task, an unsupervised learning, and then proceed with a classification task, a supervised learning. We analyze these mining tasks on both the synthetic dataset described in the above Section and on real-world and benchmark datasets.

For benchmark MTS datasets, we select 19 datasets from the UEA Multivariate Time Series Classification archive~\citep{bagnall2018uea}. Table~\ref{table:benchdata} summarizes the general description of each MTS dataset, including dataset size, time series length, number of dimensions/components UTS, number of classes, and the dataset type. The UEA Multivariate Time Series Classification archive offers a total of 30 MTS datasets; however, for simplicity, we selected the datasets with the same length and without missing values. These data are diverse, representing different types and exhibiting significant variety in terms of dimensionality. The dataset size varies between 27 and 10992, with $T \in [8, 2500]$, $m \in [2,28]$, and the number of different classes ranging from 2 to 26.

\begin{table}[hbt!]
\centering
\caption{Brief description of the benchmark multivariate time series datasets.}
\label{table:benchdata}

\begin{tabular}{|l|r|r|r|r|r|}
\hline

 \multicolumn{1}{|c|}{\multirow{2}{*}{\textbf{Dataset}}} & \multicolumn{1}{c|}{\textbf{Size of}} & \multicolumn{1}{c|}{\textbf{TS}} & \multicolumn{1}{c|}{\textbf{Num. of}} & \multicolumn{1}{c|}{\textbf{Num. of}} & \multicolumn{1}{c|}{\textbf{Type}} \\
 & \multicolumn{1}{c|}{\textbf{Dataset}} & \multicolumn{1}{c|}{\textbf{Length}} & \multicolumn{1}{c|}{\textbf{Dimens.}} & \multicolumn{1}{c|}{\textbf{Classes}} & \multicolumn{1}{c|}{\textbf{Dataset}} \\
\hline

\textbf{1-Libras} & 360 & 45 & 2 & 15 & HAR \\
\textbf{2-Epilepsy} & 275 & 206 & 3 & 4 & HAR \\
\textbf{3-Handwriting} & 1000 & 152 & 3 & 26 & HAR \\
\textbf{4-ERing} & 300 & 65 & 4 & 6 & HAR \\
\textbf{5-BasicMotions} & 80 & 100 & 6 & 4 & HAR \\
\textbf{6-RacketSports} & 303 & 30 & 6 & 4 & HAR \\
\textbf{7-ArticularyWordRecognition} & 575 & 144 & 9 & 25 & HAR \\
\textbf{8-Cricket} & 180 & 1197 & 6 & 12 & HAR \\
\textbf{9-NATOPS} & 360 & 51 & 24 & 6 & HAR \\
\textbf{10-UWaveGestureLibrary} & 4479 & 315 & 3 & 8 & HAR \\
\textbf{11-AtrialFibrillation} & 30 & 640 & 2 & 3 & ECG \\
\textbf{12-StandWalkJump} & 27 & 2500 & 4 & 3 & ECG \\
\textbf{13-SelfRegulationSCP1} & 561 & 896 & 6 & 2 & EEG \\
\textbf{14-SelfRegulationSCP2} & 380 & 1152 & 7 & 2 & EEG \\
\textbf{15-HandMovementDirection} & 234 & 400 & 10 & 4 & EEG \\
\textbf{16-FingerMovements} & 416 & 50 & 28 & 2 & EEG \\
\textbf{17-EthanolConcentration} & 524 & 1751 & 3 & 4 & OTHER \\
\textbf{18-LSST} & 4925 & 36 & 6 & 14 & OTHER \\
\textbf{19-PenDigits} & 10992 & 8 & 2 & 10 & MOTION \\

\hline
\end{tabular}
\end{table}

\subsubsection{Multivariate Time Series Clustering}

In this section, we demonstrate the utility of MNet features, particularly those related to both intra- and inter-layer edges, in MTS data mining tasks, focusing on clustering through a feature-based approach~\citep{maharaj2019time}. For a given set of MTS, we compute the MHVG feature vectors. These vectors are then Min-Max rescaled to the $[0, 1]$ range and organized in a feature data matrix. The PCs are computed without the need for z-score normalization in PCA computation. Finally, we apply the \textit{k-means} clustering algorithm to all PCs (100\% of variability). We choose \textit{k-means} for its speed and widespread use, even though it requires the pre-specification of the number of clusters, which is not a problem for this work. We employ commonly used clustering evaluation metrics: Average Silhouette (AS), Adjusted Rand Index (ARI), and Normalized Mutual Information (NMI). AS does not require ground truth, while the ARI and NMI do. The range of values is $[-1,1]$ for ARI and AS and $[0,1]$ for NMI.  

\begin{figure}[!ht]
\centering
\includegraphics[width=0.95\textwidth]{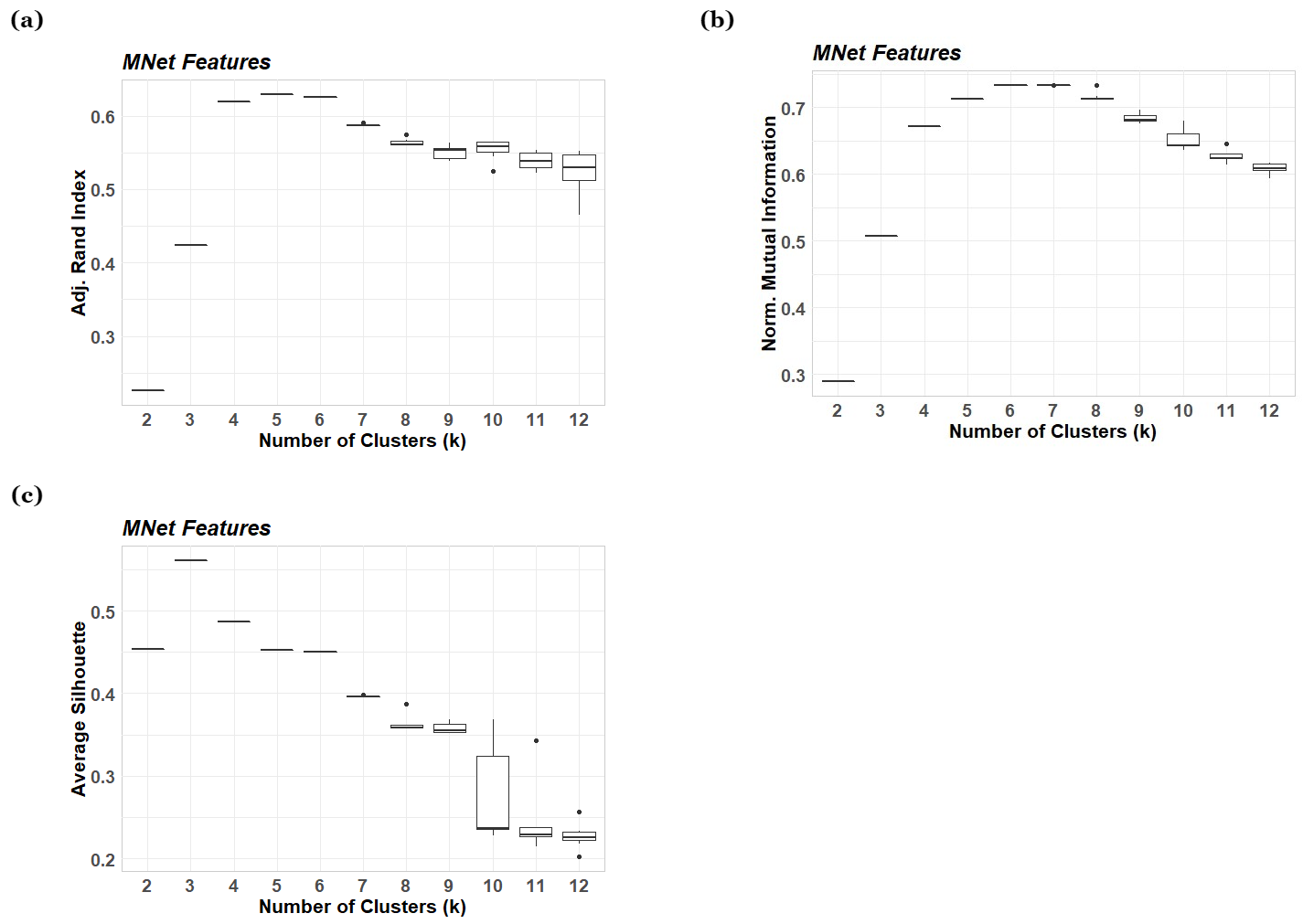}
\caption{Results of DGP's clustering evaluations using the three evaluation metrics, (a) Adjusted Rand Index (ARI), (b) Normalized Mutual Information (NMI), and (c) Average Silhouette (AS). The evaluation is computed for different numbers of clusters, $k$, given as input in the $k$-means algorithm. The results refer to 10 repetitions of the clustering analysis using the MNet features set.} \label{fig:eval_met}
\end{figure}

The clustering analysis results for the 100 instances of the six DGP datasets (see Table~\ref{tab:mts_setting}) are summarized in Figure~\ref{fig:eval_met}. This summary reflects 10 repetitions of the clustering analysis using the set of 21 MHVG features proposed in Section~\ref{sec4_3} and the $k$-means algorithm for different values of $k$ (with $k \in [2,12]$). Different evaluation metrics suggest different optimal numbers of clusters for the dataset: ARI indicates $k=5$, followed by $k = 6$ (the ground truth value), while the NMI indicates either $k=6$ or $k=7$. Metric AS, using the silhouette method to assess cluster quality indicates  $k=3$. It is interesting to note that the elements of the three clusters are: the \texttt{sVAR} models in one cluster, the VGARCH models in another, and the WN and \texttt{wVAR} models in the third cluster, indicating that the DGPs were clustered according to correlation  (serial and cross) and volatility properties.

\begin{figure}[!ht]
\centering
\includegraphics[scale = 0.35]{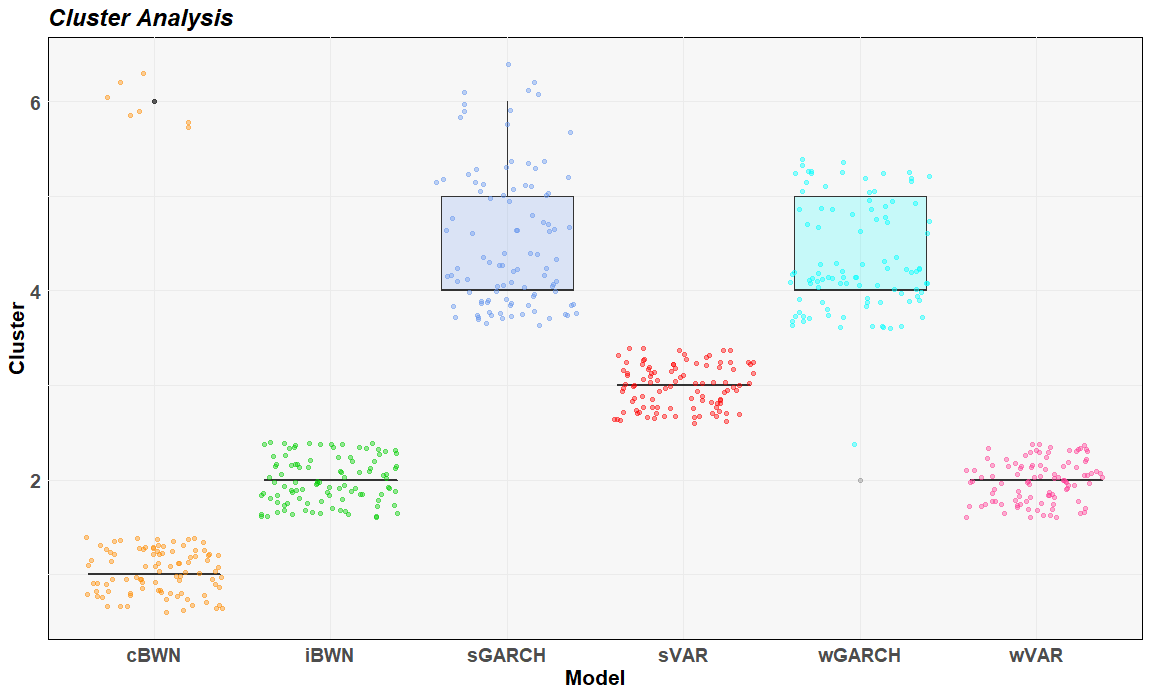}
\caption{Attribution of the samples corresponding to instances of MTS models to the different clusters based on intra, inter, all-layer, and relational feature sets. Models are displayed on the horizontal axis, each represented by a unique color. Colored points represent bivariate time series samples according to their model process, while the vertical axis denotes the assigned cluster number.} \label{fig:clust}
\end{figure}

Figure~\ref{fig:clust} shows the results for DGP clustering using the 21 MHVG features and the $k$-means algorithm with $k=6$ (ground truth). Notably, there is a perfect attribution of \texttt{cBWN} and \texttt{sVAR} samples across two different clusters (clusters 1 and 3), the attribution of \texttt{iBWN} and \texttt{wVAR} samples to the same cluster (cluster 2), and a uniform attribution of GARCH  samples (\texttt{wGARCH} and \texttt{sGARCH}) across two clusters (4 and 5). Additionally, some samples of \texttt{cBWN} and \texttt{sGARCH} are found in cluster 6, consistent with the chosen $k = 6$ and the inherent dissimilarity of VGARCH samples in the feature space. 

\newpage
We also conducted the clustering analysis considering the different MNet feature sets described in Section~\ref{sec4}. The summarized results in Table~\ref{table:clust_res} indicate that inter-layer edges contain valuable information about MTS data, leading to improved cluster outcomes. Subgraphs incorporating both intra-layer edges and inter-layer edges contribute additional information, resulting in enhanced clustering compared to using only intra-layer features (compare the last three rows with the first two). The findings reveal that Cross-HVG of inter-layer edges capture distinct properties of MTS data, yielding improved clustering outcomes as anticipated. Also, note that ARI and NMI results from the set of intra-layer features are favorable, reflecting the shared statistical process in the DGP being analyzed for the two time series components. The inherent properties of each process are effectively captured by the HVG mapping methods.

\begin{table}[!ht]
\centering
\caption{Evaluation metrics for clustering analyses with different MHVG feature vectors. Values represent the mean of 10 repetitions for various feature vectors and the ground truth ($k=6$). The highest values are highlighted.}
\label{table:clust_res}

\begin{tabular}{|l|r|r|r|}
\hline
\multicolumn{1}{|c|}{\multirow{2}{*}{\textbf{Feature Set}}} & \multicolumn{1}{c|}{\textbf{ARI}} & \multicolumn{1}{c|}{\textbf{NMI}} & \multicolumn{1}{c|}{\textbf{AS}} \\
 & \multicolumn{1}{c|}{\scriptsize $[-1,1]$} & \multicolumn{1}{c|}{\scriptsize $[0,1]$} & \multicolumn{1}{c|}{\scriptsize $[-1,1]$} \\
  
\hline
\hline
\textbf{\textit{Intra}-layer} 					& 0.522             & 0.614             & 0.289             \\
\hline
\textbf{\textit{Inter}-layer} 					& 0.294             & 0.418             & \underline{0.508} \\
\hline
\textbf{\textit{All}-layer} 					& \textbf{0.667}    & \textbf{0.725}    & \underline{0.505} \\
\hline
\textbf{\textit{Relational}}    		        & \underline{0.575} & \underline{0.646} & \textbf{0.622}    \\
\hline
\hline
\textbf{MHVG}    		                & \textbf{0.629}    & \textbf{0.713}    & 0.452             \\
\hline
\end{tabular}
\end{table}

To complement and validate our findings, we replicated the aforementioned experiment on benchmark datasets detailed in Table~\ref{table:benchdata}. Our focus was initially on evaluating the performance of the feature set derived from MHVGs in automatically determining the optimal number of clusters ($k$) for each dataset, utilizing clustering evaluation metrics such as ARI, NMI, and AS.

\newpage
We followed the same methodology, conducting 10 repetitions of clustering analyses for each dataset. The optimal value of $k$ was determined by assessing the clustering performance using the selected clustering metric. To achieve this, we performed the clustering task for $k$ values ranging from 2 to the ground truth value plus 6. The first $k$ that yielded the highest metric value was selected (i.e., the smallest $k$ if there were multiple with the highest metric value). In Appendix~\ref{app:bestk} is presented Table~\ref{table:bestk_res} with the results, that is, with the mean values corresponding to the 10 repetitions (standard deviations were non-significant). Overall, the obtained $k$ values align closely with the ground truth for ARI and NMI metrics, except for datasets \textbf{3} and \textbf{6} --- Human Activity Recognition (HAR) datasets from smartwatches --- where $k$ values significantly deviate. Furthermore, we can observe that the AS metric produced less favorable results, indicating a low intra-cluster density in the generated clusters. This observation can be attributed to the number of features used in the experiments, as we did not employ techniques to reduce dimensionality or select more representative features. Additionally, it may be influenced by the normalization method used for the feature vector, which, in this case, was the Min-Max method.

Subsequently, we fixed the value of $k$ to the ground truth of each dataset and applied the clustering methodology, as described earlier, to different subsets of topological features extracted from the corresponding MHVGs. The results of the 10 repetitions of the experiment, using the NMI evaluation metric, are presented in Table~\ref{table:subsetbench}. 

\begin{table}[hbt!]
\centering
\caption{Evaluation of multivariate time series clustering tasks involving different clustering analyses based on distinct Multilayer Network feature vectors. Results, expressed in terms of the Normalized Mutual Information metric, represent mean values from 10 repetitions of the proposed method for the different feature vectors (each column) and the ground truth. The highest values are highlighted.}
\label{table:subsetbench}

\begin{tabular}{|c|c|c|c|c|c|}
\hline

\multicolumn{1}{|c|}{\multirow{1}{*}{\textbf{Dataset}}} & \multicolumn{1}{c|}{\multirow{1}{*}{\textbf{\textit{Intra}-layer}}} & \multicolumn{1}{c|}{\multirow{1}{*}{\textbf{\textit{Inter}-layer}}} & \multicolumn{1}{c|}{\multirow{1}{*}{\textbf{\textit{All}-layer}}} & \multicolumn{1}{c|}{\multirow{1}{*}{\textbf{\textit{Relational}}}} & \multicolumn{1}{c|}{\multirow{1}{*}{\textbf{MHVG}}} \\
\hline

\textbf{1} & 0.498 & 0.259 & 0.389 & 0.493 & \textbf{0.611} \\
\textbf{2} & \textbf{0.431} & 0.140 & 0.308 & 0.151 & 0.380  \\
\textbf{3} & 0.181 & 0.156 & 0.180 & 0.146 & \textbf{0.183}  \\
\textbf{4} & 0.554 & 0.461 & 0.551 & 0.291 & \textbf{0.616}  \\
\textbf{5} & 0.437 & 0.415 & 0.731 & \textbf{0.859} & 0.823  \\
\textbf{6} & 0.098 & 0.129 & \textbf{0.153} & 0.129 & 0.146  \\
\textbf{7} & 0.651 & 0.668 & 0.790 & 0.666 & \textbf{0.797}  \\
\textbf{8} & 0.829 & 0.682 & 0.844 & 0.865 & \textbf{0.895}  \\
\textbf{9} & 0.362 & 0.261 & 0.396 & 0.351 & \textbf{0.410}  \\
\textbf{10} & 0.292 & 0.275 & \textbf{0.371} & 0.248 & 0.363  \\
\textbf{11} & 0.047 & 0.066 & 0.007 & \textbf{0.077} & 0.033  \\
\textbf{12} & 0.057 & 0.169 & \textbf{0.210} & 0.143 & 0.174  \\
\textbf{13} & 0.030 & 0.000 & \textbf{0.063} & 0.000 & 0.001  \\
\textbf{14} & 0.001 & 0.001 & \textbf{0.003} & 0.000 & 0.000  \\
\textbf{15} & 0.014 & 0.005 & 0.022 & 0.016 & \textbf{0.031}  \\
\textbf{16} & \textbf{0.002} & 0.000 & \textbf{0.002} & 0.000 & 0.000  \\
\textbf{17} & 0.007 & 0.003 & \textbf{0.010} & 0.003 & 0.004  \\
\textbf{18} & 0.099 & 0.138 & 0.136 & 0.111 & \textbf{0.181}  \\
\textbf{19} & 0.329 & 0.185 & 0.267 & 0.237 & \textbf{0.421}  \\

\hline
\end{tabular}
\end{table}

With the benchmark dataset, it becomes evident that the features associated with inter-layer edges, introduced in this work, contribute additional information that enhances the clustering results. Notably, only dataset \textbf{2} exhibits superior results for the subset of intra-layer features. Moreover, it is worth highlighting that the least favorable outcomes are observed for ECG and EEG datasets, suggesting that the topological features proposed in this work may not be optimally suited for this particular type of data record. Conversely, the most favorable results are associated with HAR and MOTION data types, particularly for datasets where classes are linked to movement determination or recognition. Interestingly, the results seem to be independent of the dataset dimensions, including the time series length, the number of components/dimensions, and the number of classes.

\subsubsection{Multivariate Time Series Classification}

We extended our analysis of MHVG feature sets to a classification problem, specifically a supervised learning task, using benchmark multivariate time series datasets. The datasets were divided into training and testing sets following the procedure outlined in ~\cite{bagnall2018uea}. Leveraging the Random Forest ensemble learning method, we conducted ten iterations of training and predictions for each dataset. Results were evaluated using the widely accepted Accuracy metric, which computes the ratio of correct predictions to the total number of predictions.

Table~\ref{table:classfbench} presents the results obtained using different sets of MHVG features, with the values representing the mean of the 10 repetitions for each experiment. The most favorable outcomes were consistently achieved when utilizing the entire set of MHVG features, combining all topological feature subsets. Notably, these results surpassed those obtained with the intra-layer feature set. This aligns with our findings from the clustering analysis, reinforcing the significance of the proposed topological inter-layer features and validating the Cross-HVG method. Similar to the clustering analysis, performance was very good for certain datasets, particularly those related to HAR and MOTION data, while less favorable for others, namely ECG and EEG data. Additionally, there appears to be no discernible relationship with the dimensionality characteristics of the analyzed data.

In conclusion, while the overall results may not exhibit exceptional strength, we consider them good and promising, especially given the global nature of the features employed and the absence of more advanced techniques, such as feature selection, which could further enhance the results.

\begin{table}[hbt!]
\centering
\caption{Evaluation of multivariate time series classification tasks involving different classification analyses based on various MHVG topological feature vectors. Results, expressed in terms of the Accuracy metric, represent the mean values from 10 repetitions of the proposed method for the different feature vectors (each column). The highest values are highlighted.}
\label{table:classfbench}

\begin{tabular}{|c|c|c|c|c|c|}
\hline

\multicolumn{1}{|c|}{\multirow{1}{*}{\textbf{Dataset}}} & \multicolumn{1}{c|}{\multirow{1}{*}{\textbf{\textit{Intra}-layer}}} & \multicolumn{1}{c|}{\multirow{1}{*}{\textbf{\textit{Inter}-layer}}} & \multicolumn{1}{c|}{\multirow{1}{*}{\textbf{\textit{All}-layer}}} & \multicolumn{1}{c|}{\multirow{1}{*}{\textbf{\textit{Relational}}}} & \multicolumn{1}{c|}{\multirow{1}{*}{\textbf{MHVG}}} \\
\hline

\textbf{1} & 0.650 & 0.257 & 0.427 & 0.588 & \textbf{0.772} \\
\textbf{2} & 0.796 & 0.662 & 0.741 & 0.636 & \textbf{0.851} \\
\textbf{3} & 0.114 & 0.096 & 0.125 & 0.068 & \textbf{0.133} \\
\textbf{4} & 0.657 & 0.611 & 0.637 & 0.613 & \textbf{0.742} \\
\textbf{5} & 0.870 & 0.792 & 0.925 & 0.887 & \textbf{0.960} \\
\textbf{6} & 0.631 & 0.631 & 0.637 & 0.603 & \textbf{0.737} \\
\textbf{7} & 0.803 & 0.846 & 0.889 & 0.816 & \textbf{0.930} \\
\textbf{8} & 0.936 & 0.825 & 0.897 & 0.936 & \textbf{0.985} \\
\textbf{9} & 0.653 & 0.647 & 0.696 & 0.681 & \textbf{0.714} \\
\textbf{10} & 0.499 & 0.426 & 0.541 & 0.509 & \textbf{0.612} \\
\textbf{11} & 0.200 & 0.253 & 0.267 & \textbf{0.333} & 0.260 \\
\textbf{12} & \textbf{0.393} & 0.387 & 0.380 & 0.333 & \textbf{0.393} \\
\textbf{13} & 0.534 & 0.594 & 0.636 & 0.553 & \textbf{0.643} \\
\textbf{14} & 0.503 & 0.538 & \textbf{0.558} & 0.491 & 0.511 \\
\textbf{15} & 0.230 & 0.269 & 0.254 & 0.227 & \textbf{0.297} \\
\textbf{16} & \textbf{0.540} & 0.479 & 0.404 & 0.476 & 0.445 \\
\textbf{17} & 0.242 & 0.235 & 0.229 & 0.249 & \textbf{0.264} \\
\textbf{18} & 0.383 & 0.428 & 0.447 & 0.419 & \textbf{0.464} \\
\textbf{19} & 0.750 & 0.581 & 0.638 & 0.748 & \textbf{0.836} \\

\hline
\end{tabular}
\end{table}

\newpage
\section{Conclusion}\label{sec6}

In this paper, we introduce a novel mapping method for representing multivariate time series as multilayer networks. Our approach involves mapping each time series component into a layer using the horizontal visibility concept and establishing inter-layer edges based on a new concept called cross-horizontal visibility. While our focus is on horizontal visibility due to promising results in prior works~\citep{vanessa2018time,vanessa2022}, the proposed Cross-HVG algorithm can naturally extend to other visibility concepts and versions of visibility graphs.

To evaluate our proposed method, we analyzed a specific set of topological features in multilayer networks, drawing inspiration from concepts such as node centrality, graph distances, clustering, communities, and similarity measures. These features are extracted from three types of subgraphs within the resulting multilayer network structure: intra-layer graphs (containing only intra-layer edges), inter-layer graphs (containing only inter-layer edges), and all-graphs (containing both intra and inter-layer edges).

We conduct an empirical evaluation on a set of 600 synthetic bivariate time series, grouped into 6 different and specific statistical models, resulting in a dataset of 600 MHVGs. To understand the potential of our proposed mapping method, we first analyze the degree distributions of the intra-, inter-, and all-layer subgraphs within MHVGs. We were able to identify the specific properties of multivariate time series models. Notably, we establish connections between weak and strong cross-correlation and the shapes of inter-layer degree distribution curves,  as well as weak and strong autocorrelation and the shapes of intra-layer degree distribution curves. Specifically, we identified that the persistence of strong correlations is related to distributions with a positively skewed shape, indicating a longer right tail. Additionally, beyond correlation properties --- both auto and cross, contemporaneous and lagged --- the properties of statistical models, such as heteroscedasticity and smoothness, result in inter-layer and all-layer degree distributions with distinct shape curves.

We also investigated the global topological features of the subgraphs (intra, inter, and all-layer). Community-related features from intra and all-layer graphs highlight the strong VAR models, with high and persistent autocorrelation and cross-correlation, as well as with smoothness, and from inter-layer graphs highlight the heteroscedasticity models, both weak and strong VGARCH models. However, the values of average path length from all-layer graphs seem to distinguish the properties of weakly and strongly correlated. The average intra-degree has higher values for independent white noise and weak VAR models but does not distinguish them.

The new relational feature proposed in this work, average ratio degree, seems to differentiate well highly correlated contemporary white noise models, resulting from the similarity between its inter-layer degree and intra-layer degree features. In the context of this work, based on the synthetic models chosen for analysis, the Jensen–Shannon divergence (JSD) measure is only useful to characterize strong VAR models that present very strong and very persistent correlations, unlike the other models. However, this feature can prove to be particularly valuable in real scenarios, where the different variables within a multivariate time series can follow different dynamic models - captured effectively by this feature. We observe this fact, for example, in some real data sets, namely, \textbf{AtrialFibrillation} and \textbf{StandWalkJump} (ECG type data), and \textbf{Epilepsy} and \textbf{Cricket} (HAR type data). Notably, in our analysis, we verify that these relational features emerge as one of the topological features that contribute most to the principal components in the resulting PCA for these datasets. 

\newpage
Our focus is not on competition or comparison with existing methods, proof of this is the non-introduction of weights on visibility edges --- a factor known to improve clustering results in the univariate setting. Instead, we emphasize the validity of our proposed mapping method and features by conducting clustering and classification mining analyses on synthetic time series and benchmark datasets. The results highlight the value of incorporating inter-layer edges within the MHVGs. This complements intra-layer edges, enabling more effective differentiation among the different statistical processes underlying the multivariate data models and improves the clustering and classification results in predicting the ground truth classes for real datasets.

While we acknowledge the potential for further advancements by integrating machine and deep learning approaches with our proposed methodology, our current focus lies in presenting an innovative approach for multivariate time series analysis based, essentially, on interpretable features. This is a crucial aspect of meaningful data analysis --- an attribute often overlooked in machine and deep learning models.

To conclude, this work introduces a procedure based on multivariate mapping to derive a set of features for multivariate time series applicable to various mining tasks. 
The method revolves around the concept of visibility, where the construction of the graph involves mapping point values onto individual nodes. While scalability challenges may arise for large datasets, as indicated by the complexity analysis conducted, future efforts could enhance computational efficiency through algorithmic techniques. Additionally, exploring alternative mapping methods, like those based on transition and proximity concepts, may also offer solutions for scalability concerns.

Furthermore, the visibility criteria addressed in this work may exhibit sensitivity, implying less resistance to noise present in real data. In future work, we plan to analyze our methodology within the Limited Penetrable Visibility Graph~\citep{ning2012limited} method, as it demonstrated noise resistance in the univariate setting.

Looking ahead, our future work will explore bridging the aforementioned gap by not only integrating our methodologies with advanced machine and deep learning techniques but also by conducting comparative analyses with recent literature. Additional open issues include exploring new topological features of multilayer networks and incorporating advanced feature selection techniques (supervised and/or unsupervised) to reduce the number of topological features that increase with the number of time series components. 
 
This comprehensive exploration aims to position our work within the evolving landscape of time series data analysis, offering both interpretable features and the potential for further performance enhancements.

\section*{Availability of data and materials}
The raw data is available at \url{https://github.com/vanessa-silva/MHVG2MTS}.

\section{Acknowledgments}
We want to thank the reviewers for the valuable comments that improved the paper. 
National Funds partially fund this work through the Portuguese funding agency, FCT -- Fundação para a Ciência e a Tecnologia, within project LA/P/0063/2020. DOI 10.54499/LA/P/0063/2020 | \url{https://doi.org/10.54499/LA/P/0063/2020}

\newpage

\bibliographystyle{apalike}
\bibliography{refs}

\newpage

\appendix

\section{Cross-HVG Algorithm}\label{app:mhvg}

See Algorithm~\ref{alg:chvg}.
\begin{algorithm}[!ht]
\scriptsize
    \caption{Cross-Horizontal Visibility Graph}\label{alg:chvg}

    \hspace*{\algorithmicindent} \textbf{Input:} {two rescaled time series ($\boldsymbol{tsA}$, $\boldsymbol{tsB}$), associated maximum time series ($\boldsymbol{tsMax}$), horizontal visibility graph layers ($\boldsymbol{layerA}$, $\boldsymbol{layerB}$), and multilayer network ($\boldsymbol{mnet}$)}
    \\
    \hspace*{\algorithmicindent} \textbf{Output:} {inter-layer edges between two layers ($layerA, layerB$) of $mnet$}

    \begin{algorithmic}[1]

    \Procedure{CHVG}{$tsA, tsB, tsMax, layerA, layerB, mnet$} 
        \State $T \gets tsMax.size()$ 
        
        \For{$i \gets 1$ \textbf{to} $T-1$}
            \State $k \gets i+1$
            \For{$j \gets i+1$ \textbf{to} $T$}
                \If{$j == i+1$} 
                    \State $mnet.$\Call{add\_Edge}{$i, j, layerA, layerB$}
                \Else 
                    \If{$tsMax[k] >= tsA[i]$}
                        \State $break$
                    \EndIf
                    \If{$tsMax[k] < tsB[j]$}
                        \State $mnet.$\Call{add\_Edge}{$i, j, layerA, layerB$}
                        \State $k \gets j$
                    \Else
                        \If{$tsMax[j] > tsMax[k]$}
                            \State $k \gets j$
                        \EndIf
                    \EndIf
                \EndIf
            \EndFor
        \EndFor

        \For{$i \gets 2$ \textbf{to} $T$}
            \State $k \gets i-1$
            \For{$j \gets i-1$ \textbf{to} $1$}
                \If{$j == i-1$} 
                    \State $mnet.$\Call{add\_Edge}{$i, j, layerA, layerB$}
                \Else 
                    \If{$tsMax[k] >= tsA[i]$}
                        \State $break$
                    \EndIf
                    \If{$tsMax[k] < tsB[j]$}
                        \State $mnet.$\Call{add\_Edge}{$i, j, layerA, layerB$}
                        \State $k \gets j$
                    \Else
                        \If{$tsMax[j] > tsMax[k]$}
                            \State $k \gets j$
                        \EndIf
                    \EndIf
                \EndIf
            \EndFor
        \EndFor

        \State \Return{}

    \EndProcedure

    \end{algorithmic}
\end{algorithm}

\newpage

\section{Multivariate Time Series Models}\label{app:mts_models}

Main references~\citep{cipra2020time, wei2018multivariate, Sumway2017, tsay2013multivariate}.

\medskip
\noindent\textit{\textbf{Linear Models}}

\begin{description} 
    \item[\textbf{BWN}] The \textit{vector white noise} process, $\boldsymbol{\epsilon}_t$, is a vector of sequences of i.i.d. random variables with mean vector 0 and covariance matrix function $\boldsymbol{\Sigma}$, where $\boldsymbol{\Sigma}$ is an $m \times m$ symmetric positive definite matrix. 
    The components of the white noise process are serially uncorrelated $\mbox{\rm corr}(\epsilon_{i,t}, \epsilon_{i,s})=0$, for $t \neq s$, but may be contemporaneously correlated, $\mbox{\rm corr}(\epsilon_{i,t}, \epsilon_{j,t}) \neq 0$. 
    It is the simplest multivariate time series process that reflects information that is neither directly observable nor predictable.
    We generate white noise processes, $\boldsymbol{\epsilon}_{t} \sim N(0,1)$, that are not correlated, that is, are independent, and we refer to theses processes as \texttt{iBWN}, 
    and white noise processes contemporaneously correlated that we refer to them as \texttt{cBWN}, $\bigl[ \begin{smallmatrix} \epsilon_{1,t} \\ \epsilon_{2,t} \end{smallmatrix} \bigr] \sim N\left(0, \bigl[ \begin{smallmatrix} 1.00 & 0.86\\ 0.86 & 1.50 \end{smallmatrix} \bigr] \right)$. 
    \vspace{2.5mm}
    
    \item[\textbf{VAR$(1)$}] The \textit{vector autoregression} process is a natural extension of the univariate \textit{autoregressive} (AR) process that the variable values depend linearly on its previous values and a stochastic term. 
    We defined a VAR$(1)$ process as a vector AR process of order 1 if it satisfies the following equation: 
        \begin{equation} \label{eq_var}
	        \boldsymbol{Y}_{t} = \boldsymbol{\varphi} + \boldsymbol{\phi} \boldsymbol{Y}_{t-1} + \boldsymbol{\epsilon}_{t}, 
        \end{equation}
    where $\boldsymbol{\epsilon}_{t}$ is the vector white noise, $\boldsymbol{\phi}$ is the vector of autoregressive constants and $\boldsymbol{\varphi}$ is the vector of intercepts. 
    We generate a VAR$(1)$ of 2 dimensions as follows:
        \begin{equation} \label{eq_var2}
	        \bigl[ \begin{smallmatrix} Y_{1,t} \\ Y_{2,t} \end{smallmatrix} \bigr] =
	        \bigl[ \begin{smallmatrix} \varphi_{1,1} \\ \varphi_{2,1} \end{smallmatrix} \bigr] +
	        \bigl[ \begin{smallmatrix} \phi_{1,1} & \phi_{1,2} \\ \phi_{2,1} & \phi_{2,2} \end{smallmatrix} \bigr] \bigl[ \begin{smallmatrix} Y_{1,t-1} \\ Y_{2,t-1} \end{smallmatrix} \bigr] +
	        \bigl[ \begin{smallmatrix} \epsilon_{1,t} \\ \epsilon_{2,t} \end{smallmatrix} \bigr],
        \end{equation}
    where $\boldsymbol{\varphi} = \bigl[ \begin{smallmatrix} 2.50 \\ 0.50 \end{smallmatrix} \bigr]$, $\boldsymbol{\phi} = \bigl[ \begin{smallmatrix} 0.20 & 0.10 \\ 0.02 & 0.10 \end{smallmatrix} \bigr] $ and $\boldsymbol{\epsilon}_t \sim \bigl[ \begin{smallmatrix} 1.00 & 0.10 \\ 0.10 & 1.50 \end{smallmatrix} \bigr]$ to generate weakly correlated VAR$(1)$ processes, and 
    $\boldsymbol{\varphi} = \bigl[ \begin{smallmatrix} 0 \\ 0 \end{smallmatrix} \bigr]$,  $\boldsymbol{\phi} = \bigl[ \begin{smallmatrix} 0.70 & 0.02 \\ 0.30 & 0.80 \end{smallmatrix} \bigr] $ and $\boldsymbol{\epsilon}_t \sim \bigl[ \begin{smallmatrix} 1.00 & 0.86 \\ 0.86 & 1.50 \end{smallmatrix} \bigr]$ to generate strongly correlated VAR$(1)$ processes. 
	We refer to the two models generated as \texttt{wVAR} and \texttt{sVAR}, respectively.
    
\end{description}

\medskip
\noindent\textit{\textbf{Non Linear Models}}

\begin{description}
    \item[\textbf{VGARCH$(1,1)$}] Also \textit{generalized autoregressive conditional heteroskedasticity} (GARCH) models can be generalized to multidimensional settings, extending the principle of univariate conditional heteroscedasticity to mutual volatility. 
    We generate a bivariate GARCH$(1,1)$ model according to the following volatility equation:
        \begin{equation} \label{eq_vgarch}
	        \boldsymbol{\sigma}_{t} = \boldsymbol{\omega} + \boldsymbol{\alpha} \boldsymbol{\epsilon^2}_{t-1} + \boldsymbol{\beta}\boldsymbol{\sigma}_{t-1},
        \end{equation}
    where $\boldsymbol{\sigma}_t$ denotes the volatility in the variables $\boldsymbol{Y}_t$. 
    We generate a VGARCH$(1,1)$ of 2 dimensions with the following vector of parameters:
        \begin{equation} \label{eq_vgarch2}
	        \bigl[ \begin{smallmatrix} \sigma_{11,t} \\ \sigma_{22,t} \end{smallmatrix} \bigr] =
	        \bigl[ \begin{smallmatrix} \omega_{1,1} \\ \omega_{2,1} \end{smallmatrix} \bigr] +
	        \bigl[ \begin{smallmatrix} \alpha_{1,1} & \alpha_{1,2} \\ \alpha_{2,1} & \alpha_{2,2} \end{smallmatrix} \bigr] \bigl[ \begin{smallmatrix} \epsilon^2_{1,t-1} \\ \epsilon^2_{2,t-1} \end{smallmatrix} \bigr] +
	        \bigl[ \begin{smallmatrix} \beta_{1,1} & \beta_{1,2} \\ \beta_{2,1} & \beta_{2,2} \end{smallmatrix} \bigr] \bigl[ \begin{smallmatrix} \sigma_{11,t} \\ \sigma_{22,t} \end{smallmatrix} \bigr],
        \end{equation}
    where $\boldsymbol{\epsilon}_t \sim \bigl[ \begin{smallmatrix} 1.00 & 0.10 \\ 0.10 & 1.50 \end{smallmatrix} \bigr]$ to generate weakly correlated VGARCH$(1,1)$ processes, and
    $\boldsymbol{\epsilon}_t \sim \bigl[ \begin{smallmatrix} 1.00 & 0.86 \\ 0.86 & 1.50 \end{smallmatrix} \bigr]$ to generate strongly correlated VGARCH$(1,1)$ processes. To both processes we use $\boldsymbol{\omega} = \bigl[ \begin{smallmatrix} 0.05 \\ 0.02 \end{smallmatrix} \bigr]$, $\boldsymbol{\alpha} = \bigl[ \begin{smallmatrix} 0.10 & 0.00 \\ 0.00 & 0.05 \end{smallmatrix} \bigr] $ and $\boldsymbol{\beta} = \bigl[ \begin{smallmatrix} 0.85 & 0.00 \\ 0.00 & 0.88 \end{smallmatrix} \bigr] $. 
	We refer to the two models generated as \texttt{wGARCH} and \texttt{sGARCH}, respectively.
    \vspace{2.5mm}

\end{description}

Bivariate time series are generated from the above DGP using the R packages:  \texttt{lgarch}~\cite{lgarch}, \texttt{mAr}~\cite{mAr} and \texttt{ccgarch}~\cite{ccgarch}.

\begin{figure}[ht]
  \centering
  \includegraphics[scale = 0.55]{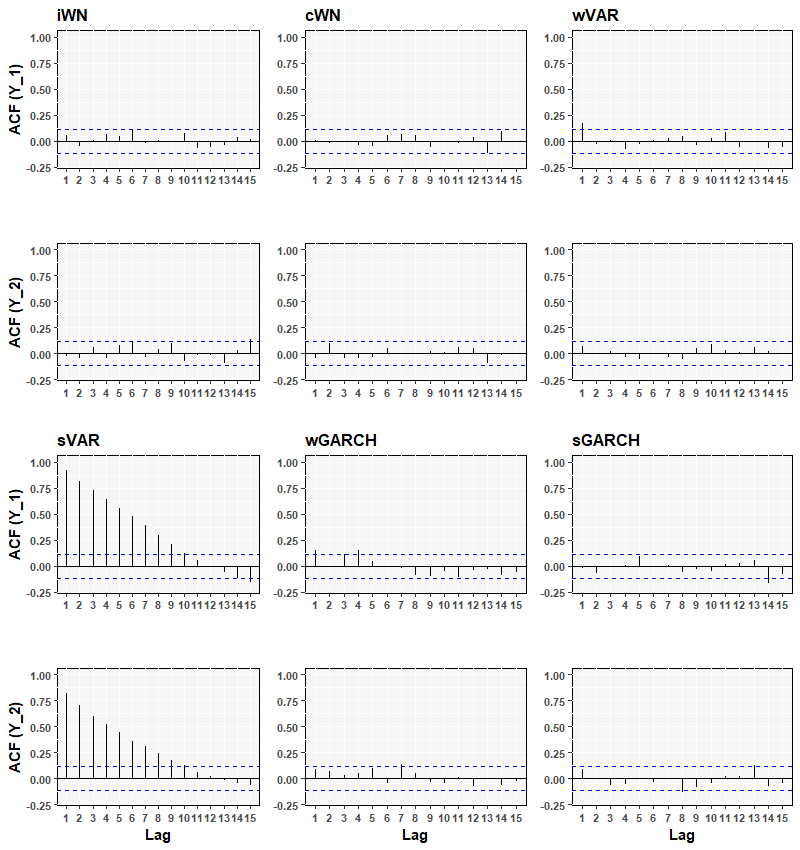}
  \caption{Plot of the autocorrelation function of an instance of each of the DGP models. 
  The first column refers to the ACFs of the first time series component ($\boldsymbol{Y}_{1}$) of each model, while the second column refers to the second component ($\boldsymbol{Y}_{2}$).}
  \label{fig:acf}
\end{figure}

Figure~\ref{fig:acf} represents the Autocorrelation Function Plots for one instance of each bivariate time series model generated.

\newpage
\subsection{Principal Component Analysis Results}

\begin{figure}[ht]
  \centering
  \includegraphics[width=1\textwidth]{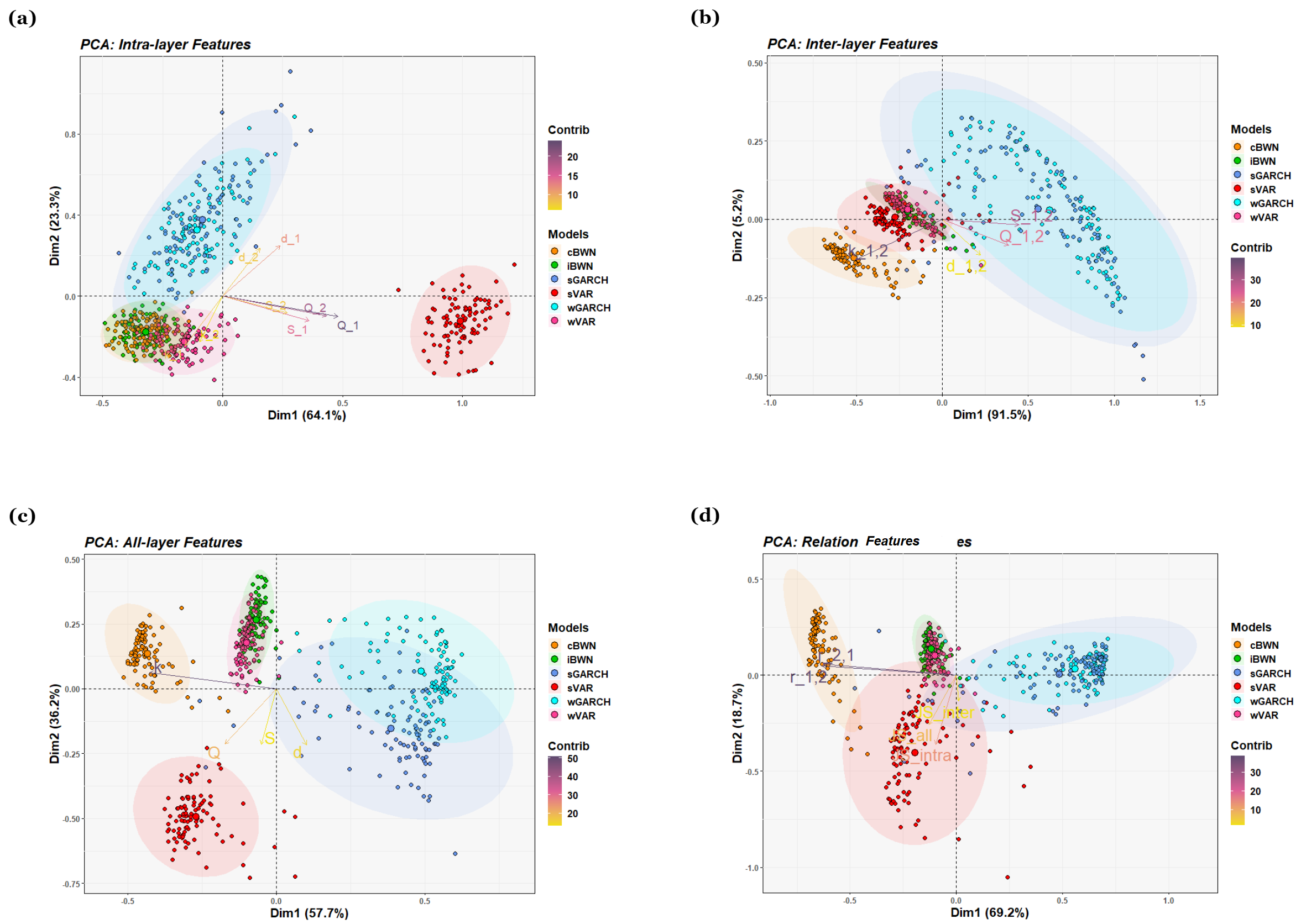}
  \caption{Bi-plot of the first two principal components (PC) of principal component analysis for the Data Generating Process (DGP) using the different feature vectors. Each DGP is represented by a different color and the arrows represent the contributions of the set of features to the PC's, the larger the size, sharpness, and closer to the red the greater the contribution of the feature.}
  \label{fig:pcaplot_sets}
\end{figure}

\begin{figure}[ht]
  \centering
  \includegraphics[scale = 0.31]{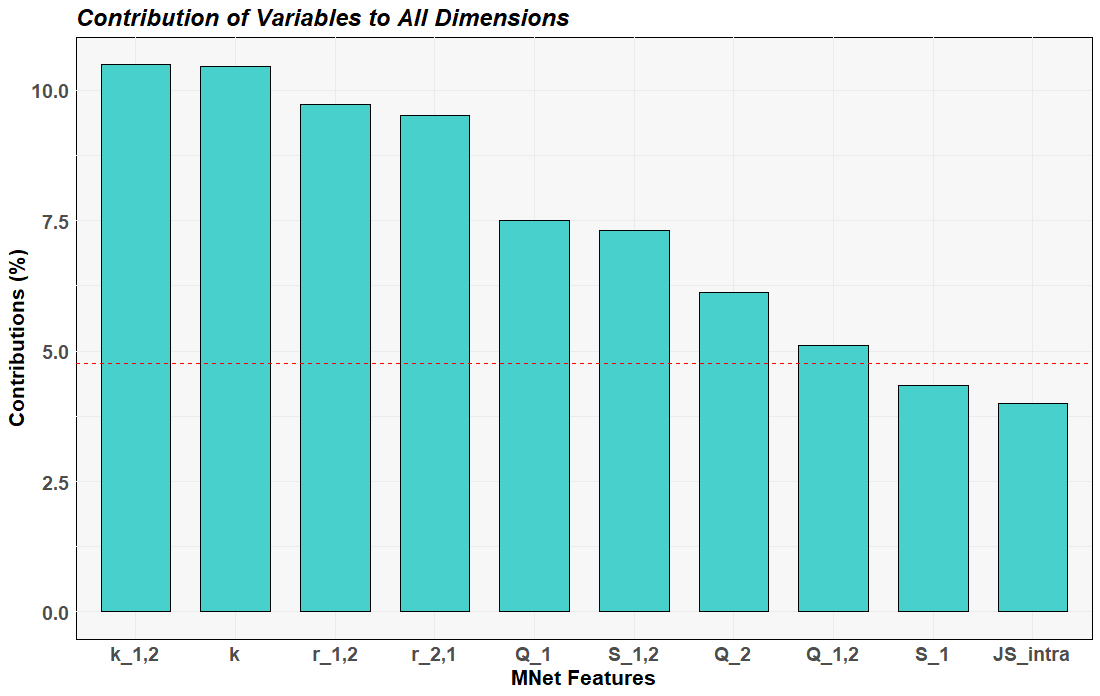}
  \caption[Bar plot with contributions of 20 MNet features]{Bar plot with contributions of MNet features to the total of all 20 principal components formed by the PCA. 
  The red dashed line on the plot indicates the expected average contribution.}
  \label{fig:barplot}
\end{figure}

See Figure~\ref{fig:pcaplot_sets} and Figure~\ref{fig:barplot}.

\newpage

\section[Automatic Determination of number of clusters using MHVG Features]{Automatic Determination of $k$ using MHVG Features}\label{app:bestk}

\begin{table}[hbt!]
\centering
\caption{Evaluation of multivariate time series clustering tasks based on the MHVG feature set, incorporating automatic determination of the number of clusters ($k$), using evaluation metrics. 
The second column provides the ground truth for each benchmark dataset for comparison, and the subsequent columns showcase the automatically determined optimal number of clusters $k$ and the corresponding evaluation metric results.}
\label{table:bestk_res}

\begin{tabular}{|c|c|c|c|c|c|c|c|}
\hline

\multicolumn{1}{|c|}{\multirow{2}{*}{\textbf{Dataset}}} & \multicolumn{1}{c|}{\textbf{Ground}} & \multicolumn{2}{c|}{\textbf{ARI}} & \multicolumn{2}{c|}{\textbf{NMI}} & \multicolumn{2}{c|}{\textbf{AS}} \\
    & \multicolumn{1}{c|}{\textbf{Truth}} & \multicolumn{1}{c}{\footnotesize $[-1,1]$} & \multicolumn{1}{c|}{\footnotesize $k$} & \multicolumn{1}{c}{\footnotesize $[0,1]$} & \multicolumn{1}{c|}{\footnotesize $k$} & \multicolumn{1}{c}{\footnotesize $[-1,1]$} & \multicolumn{1}{c|}{\footnotesize $k$} \\
\hline

\textbf{1} & 15 &  \multicolumn{1}{c}{0.366} & 15 & \multicolumn{1}{c}{0.616} & 16 & \multicolumn{1}{c}{0.261} & 4 \\
\textbf{2} & 4 & \multicolumn{1}{c}{0.377} & 6 & \multicolumn{1}{c}{0.395} & 6 & \multicolumn{1}{c}{0.160} & 2 \\
\textbf{3} & 26 & \multicolumn{1}{c}{0.026} & 14 & \multicolumn{1}{c}{0.198} & 31 & \multicolumn{1}{c}{0.288} & 2 \\
\textbf{4} & 6 & \multicolumn{1}{c}{0.528} & 7 & \multicolumn{1}{c}{0.616} & 6 & \multicolumn{1}{c}{0.192} & 2 \\
\textbf{5} & 4 & \multicolumn{1}{c}{0.816} & 4 & \multicolumn{1}{c}{0.823} & 4 & \multicolumn{1}{c}{0.227} & 2 \\
\textbf{6} & 4 & \multicolumn{1}{c}{0.196} & 9 & \multicolumn{1}{c}{0.279} & 9 & \multicolumn{1}{c}{0.093} & 2 \\
\textbf{7} & 25 & \multicolumn{1}{c}{0.644} & 30 & \multicolumn{1}{c}{0.812} & 27 & \multicolumn{1}{c}{0.152} & 2 \\
\textbf{8} & 12 & \multicolumn{1}{c}{0.832} & 15 & \multicolumn{1}{c}{0.890} & 13 & \multicolumn{1}{c}{0.197} & 9 \\
\textbf{9} & 6 & \multicolumn{1}{c}{0.322} & 9 & \multicolumn{1}{c}{0.434} & 7 & \multicolumn{1}{c}{0.096} & 3 \\
\textbf{10} & 8 & \multicolumn{1}{c}{0.233} & 6 & \multicolumn{1}{c}{0.362} & 9 & \multicolumn{1}{c}{0.271} & 2 \\
\textbf{11} & 3 & \multicolumn{1}{c}{0.005} & 5 & \multicolumn{1}{c}{0.166} & 6 & \multicolumn{1}{c}{0.214} & 3 \\
\textbf{12} & 3 & \multicolumn{1}{c}{0.108} & 2 & \multicolumn{1}{c}{0.244} & 8 & \multicolumn{1}{c}{0.140} & 4 \\
\textbf{13} & 2 & \multicolumn{1}{c}{0.036} & 4 & \multicolumn{1}{c}{0.039} & 4 & \multicolumn{1}{c}{0.235} & 2 \\
\textbf{14} & 2 & \multicolumn{1}{c}{-0.002} & 2 & \multicolumn{1}{c}{0.003} & 6 & \multicolumn{1}{c}{0.177} & 2 \\
\textbf{15} & 4 & \multicolumn{1}{c}{0.018} & 4 & \multicolumn{1}{c}{0.031} & 4 & \multicolumn{1}{c}{0.067} & 2 \\
\textbf{16} & 2 & \multicolumn{1}{c}{0.001} & 6 & \multicolumn{1}{c}{0.004} & 6 & \multicolumn{1}{c}{0.090} & 2 \\
\textbf{17} & 4 & \multicolumn{1}{c}{0.007} & 7 & \multicolumn{1}{c}{0.012} & 7 & \multicolumn{1}{c}{0.415} & 2 \\
\textbf{18} & 14 & \multicolumn{1}{c}{0.103} & 4 & \multicolumn{1}{c}{0.173} & 10 & \multicolumn{1}{c}{0.120} & 3 \\
\textbf{19} & 10 & \multicolumn{1}{c}{0.302} & 7 & \multicolumn{1}{c}{0.416} & 11 & \multicolumn{1}{c}{0.237} & 2 \\

\hline
\end{tabular}
\end{table}

\end{document}